\documentclass[iop]{emulateapj}
\usepackage{natbib}
\usepackage{longtable}
\usepackage{xcolor}
\newcommand{\kms}{\ensuremath{{\rm km}\,{\rm s}^{-1}}}
\newcommand{\vrot}{$v_{\rm rot}\sin{i}$}

\begin{document}
\title{EVR-CB-004: An Inflated Hot Subdwarf O star + Unseen WD Companion\\ in a Compact Binary Discovered with the Evryscope }

\author{Jeffrey~K.~Ratzloff\altaffilmark{1}, Thomas~Kupfer\altaffilmark{2}, Brad~N.~Barlow\altaffilmark{3}, David~Schneider\altaffilmark{4}, Thomas~R.~Marsh\altaffilmark{5},\\ Ulrich~Heber\altaffilmark{4}, Kyle~A.~Corcoran\altaffilmark{3,6}, Evan Bauer\altaffilmark{2}, Steven H\"{a}mmerich\altaffilmark{4}, Henry~T.~Corbett\altaffilmark{1}, Amy~Glazier\altaffilmark{1},\\  Ward~S.~Howard\altaffilmark{1}, Nicholas~M.~Law\altaffilmark{1}}

\altaffiltext{1}{Department of Physics and Astronomy, University of North Carolina at Chapel Hill, Chapel Hill, NC 27599-3255, USA}
\altaffiltext{2}{Kavli Institute for Theoretical Physics, University of California Santa Barbara, Santa Barbara, CA 93106, USA}
\altaffiltext{3}{Department of Physics and Astronomy, High Point University, High Point, NC 27268, USA}
\altaffiltext{4}{Dr. Karl Remeis-Observatory \& ECAP, Astronomical Institute, Friedrich-Alexander University Erlangen-Nuremberg, Sternwartstr.
7, 96049 Bamberg, Germany}
\altaffiltext{5}{Department of Physics, University of Warwick, Coventry, CV4 7AL, UK}
\altaffiltext{6}{Department of Astronomy, University of Virginia, Charlottesville, VA 22904, USA}

\email[$\star$~E-mail:~]{jeff215@live.unc.edu}


\begin{abstract}
We present the discovery of EVR-CB-004, a close binary with a remnant stellar core and an unseen white dwarf companion. The analysis in this work reveals the primary is potentially an inflated hot subdwarf (sdO) and more likely is a rarer post-blue horizontal branch (post-BHB) star. Post-BHBs are the short-lived shell-burning final stage of a blue horizontal star or hot subdwarf before transitioning to a WD. This object was discovered using Evryscope photometric data in a southern-all-sky hot subdwarf variability survey. The photometric light curve for EVR-CB-004 shows multi-component variability from ellipsoidal deformation of the primary and from Doppler boosting as well as gravitational limb darkening. EVR-CB-004 is one of just a handful of known systems, and has a long period (6.08426 hours) and large amplitude ellipsoidal modulation (16.0 \% change in brightness from maximum to minimum) for these extremely close binary systems, while the properties of the primary make it a truly unique system.  EVR-CB-004 also shows a peculiar low-amplitude (less than 1\%) sinusoidal light curve variation with a period that is a 1/3 resonance of the binary period. We tentatively identify this additional variation source as a tidally-induced resonant pulsation, and we suggest followup observations that could verify this interpretation. From the evolutionary state of the system, its components, and its mass fraction, EVR-CB-004 is a strong merger candidate to form a single high-mass ($\approx1.2M_{\odot}$) WD.  EVR-CB-004 offers a glimpse into a brief phase of a remnant core evolution and secondary variation, not seen before in a compact binary.\\

\end{abstract}


\section{INTRODUCTION} \label{section_intro}

Hot subdwarfs are small, dense stars, under-luminous for their high temperatures. They are divided into two main spectroscopic categories: B-type subdwarfs (sdB), which have temperatures from 20,000-40,000K, and O-type subdwarfs (sdO), which have temperatures from 40,000-100,000K (see \citealt{2009ARA&A..47..211H} for a description of hot subdwarf properties and types). SdOs tend to exhibit a wider range in their physical attributes; for a few recent examples see \cite{2017MNRAS.465.3101J}, and for a large sample of sdO atmospheric parameters see \cite{2007A&A...462..269S}. They are also rarer than their sdB counterparts, seen at an $\approx 1/3$ sdO/sdB ratio. A wide array of stars with different evolutionary histories fall within the ``hot subdwarf box,'' including extended horizontal branch (EHB) stars, pre/post-EHB stars, blue horizontal branch (BHB) stars, post-BHB stars, post-asymptotic giant branch (post-AGB) stars, and even pre-helium white dwarfs (pre-He WD). A recent review of hot subdwarfs can be found in \cite{2016PASP..128h2001H}, including a description of all formation channels. An analysis on the evolution of EHB stars, along with a helpful discussion on the potentially confusing terminology of EHB/HB/hot subdwarfs can be found in \cite{2009CoAst.159...75O}.

The majority of hot subdwarfs are compact helium core burning stars with a thin hydrogen shell, a canonical size of $R=0.2R_{\odot}$ and $M=0.5M_{\odot}$, and temperatures greater than $\approx$ 20,000K. They are thought to form through one of two main mechanisms: (i) the merging of two helium--core white dwarfs (WDs), or (ii) binary interactions involving Roche lobe overflow (RLOF) or common envelope (CE) evolution that result in significant hydrogen stripping from a red giant progenitor. We demonstrate further in the manuscript that the latter mechanism is relevant to this work and thought to occur when the hot subdwarf progenitor is near the tip of the red giant branch. The process leaves behind a binary system with a hot subdwarf and a companion including white dwarfs, red dwarfs, Solar--type stars, and, in some cases, substellar objects. Without a thick outer hydrogen layer, hot subdwarfs generally will neither ascend the asymptotic giant branch (AGB) nor experience the traditional planetary nebula phase, as expected for low--mass stars, but instead will evolve directly onto the white dwarf cooling sequence. Depending on their hydrogen envelope hot subdwarfs are considered to be extreme horizontal branch (EHB) stars (for hydrogen envelopes $\lesssim0.01$\,M$_\odot$) or blue horizontal branch (BHB) stars (for hydrogen envelopes of a few hundreds\,M$_\odot$). 

Hot subdwarf progenitor systems with comparatively smaller and closer companions are thought to be unable to accrete matter (from the hydrogen shell of the red-giant, hot subdwarf progenitor) at a fast enough rate to be stable. A CE forms and some matter is ejected from the system, resulting in a loss of angular momentum and tightening of the binary. A full description of the CE formation channel can be found in \cite{2008MmSAI..79..375H} and in \cite{2002MNRAS.336..449H, 2003MNRAS.341..669H}. Post--CE hot subdwarf binaries typically have periods from 2 hours up to 30 days, with a few known exceptionally short period systems. Common companions are M-dwarfs, K-dwarfs, and white dwarfs; more exotic remnant companions are also possible.

The CE formation channel for sdB and sdO stars is modelled extensively in \cite{2002MNRAS.336..449H, 2003MNRAS.341..669H}, with simulations resulting in short period binaries between 2 hours and 10 days, and a hot subdwarf mass near $0.46M_{\odot}$. Different initial conditions, including the hydrogen shell mass, helium core mass, and mass loss, lead to a range of temperature and surface gravity values that are in general agreement with observed sdB and sdO binaries.

A rare and interesting subset of post--CE hot subdwarf binaries are the compact, very short period binaries with unseen white dwarf (WD) companions. Only a handful of these systems are known after decades of searching. To highlight these systems: KDP 1930+2752 \citep{1986ApJS...61..569D} is a high mass system found as part of the Kitt Peak - Downes survey of UV excess objects, later determined by \cite{2000ApJ...530..441B} to be a 2.28 hour period binary sdB + WD. Work by \cite{2000MNRAS.317L..41M} identified this system to be a strong SN Ia progenitor candidate. The slightly lower mass but shorter period sdB + WD binary systems KPD 0422+5421 \citep{1998MNRAS.300..695K, 1999MNRAS.310..773O} and CD-30 11223 (first reported in \citealt{2041-8205-759-1-L25}, with subsequent followup in \citealt{2013A&A...554A..54G}) are the only systems that show evidence of eclipses, helping to separately verify the sdB radius, and to constrain the inclination angle as well as the sdB and WD sizes more tightly. PTF1J082340.04+081936.5 \citep{Kupfer_2017}, is the second shortest period system at 1.41 hours and has a low mass WD companion, a surprising find in such a tight orbit. The recent discovery of EVR-CB-001 \citep{2019ApJ...883...51R} reveals a 2.34 hour period compact binary system with exceptionally low mass components. The primary is a rare transitioning object (pre-He-WD) appearing as an sdB in color magnitude space, and the system is a strong merger candidate to form a single hot subdwarf (single hot subdwarfs are observed but their formation is difficult to explain). Lastly, OWJ074106.0-294811.0 \citep{2017ApJ...851...28K} is an ultra-compact (44.7 minute) sdO + WD system with a non-canonical mass sdO.

The photometric light curves in the above systems show sinusoidal-like variations due to ellipsoidal deformation of the hot subdwarf by the WD companion, with differences between even and odd phases due to Doppler boosting and gravity darkening. These unique light curve features, combined with spectral and radial velocity analysis, allow for precise solutions to the system. The multi-component photometric variations can aid in the discovery of these rare systems, however the detections are challenging as the half-period alias folded light curves look nearly indistinguishable from an unexceptional variable with a simple sinusoidal signal.

In this work we present the discovery of EVR-CB-004, cataloged as Gaia DR2 5642627428172190000, an sdO hot subdwarf  + WD compact binary with a 6.084 hour period. The hot subdwarf is likely a post-BHB or post-AGB star. EVR-CB-004 shows strong multi component photometric variability, high radial velocity amplitudes, and is bright ($m_{G}=13.1$), characteristics that aid in the system solution. EVR-CB-004 was found in a southern all-sky hot subdwarf survey searching for low mass companions \citep{2020ApJ...890..126R} using the Evryscope \citep{2019PASP..131g5001R, 2015PASP..127..234L}, a new type of telescope with fast-cadence and all-sky capability.

This paper is organized as follows: in \S~\ref{section_obs} we describe the discovery and observations. In \S~\ref{section_analysis_spectra} we describe our spectroscopic analysis to determine the orbital and atmospheric parameters of the sdO. In \S~\ref{section_analysis_lc} we model the photometric light curve to determine ellipsoidal modulations and test for eclipses. In \S~\ref{section_system_solution} we solve the system and show our results. In \S~\ref{section_discussion} we discuss our findings which reveal several unexpected features of the system. The sdO is shown to have a considerably lower surface gravity ($\log{g}=4.55$) than expected for a standard shell-burning sdO hot subdwarf (typically $\log{g}=5.5-6.0$ see \citealt{2009CoAst.159...75O}), with a corresponding large radius of $0.6R_{\odot}$. The primary in EVR-CB-004 is likely a more evolved hot subdwarf, found during the final stage (known as a post-BHB) of its evolution before forming a WD - a surprising find in an already rare compact binary system. In addition to the ellipsoidal modulation, Doppler boosting, gravity darkening and limb darkening components, the light curve of EVR-CB-004 also shows a sinusoidal variation at the $0.4\%$ level with a period that is a 1/3rd resonance (2.028 hours) of the orbital period. We identify this as a pulsation, with the suggestion of further followup work to confirm. We conclude in \S~\ref{section_summary}.


\section{OBSERVATIONS AND REDUCTION} \label{section_obs}

\subsection{Evryscope Photometry}

Evryscope photometric observations taken from February 2017 to June 2017 led to the discovery of EVR-CB-004. Data were taken through a Sloan {\em g} filter with 120 s integration times, providing a total of 4,812 measurements. The wide-seeing Evryscope is a gigapixel-scale, all-sky observing telescope that provides new opportunities for uncovering rare compact binaries through photometric variations. It is optimized for short-timescale observations with continuous all sky coverage and multiple years of observations for all targets. The Evryscope is a robotic camera array mounted into a 6 ft-diameter hemisphere which tracks the sky \citep{2015PASP..127..234L, 2019PASP..131g5001R}. The instrument is located at CTIO in Chile and observes continuously, covering 8150 sq. deg. in each 120s exposure. Each camera features a 29MPix CCD providing a plate scale of 13"/pixel. The Evryscope monitors the entire accessible Southern sky at 2-minute cadence, and the Evryscope database includes tens of thousands of epochs on 16 million sources.

The Evryscope EVR-CB-004 light curve has a less than average number of data points because observations for additional seasons (the Evryscope has been observing since mid 2015) were removed as problematic points due to the difficult observing field (source crowding and unfavorable airmass). The additional epochs were not necessary for the discovery of EVR-CB-004, but are expected to be recovered with the upgraded photometric pipeline (currently processing light curves for all Evryscope sources including 2019 observations).

Here we only briefly describe the calibration, reduction, and extraction of light curves from the Evryscope; for further details we point the reader to our Evryscope instrumentation paper \citep{2019PASP..131g5001R}. Raw images are filtered with a quality check, calibrated with master flats and master darks, and have large-scale backgrounds removed using the custom Evryscope pipeline. Forced photometry is performed using APASS-DR9 \citep{2015AAS...22533616H} as our master reference catalog. Aperture photometry is performed on all sources using multiple aperture sizes; the final aperture for each source is chosen to minimize light curve scatter. Systematics removal is performed with a custom implementation of the SysRem \citep{2005MNRAS.356.1466T} algorithm.

\begin{figure}[ht]
\includegraphics[width=1.0\columnwidth]{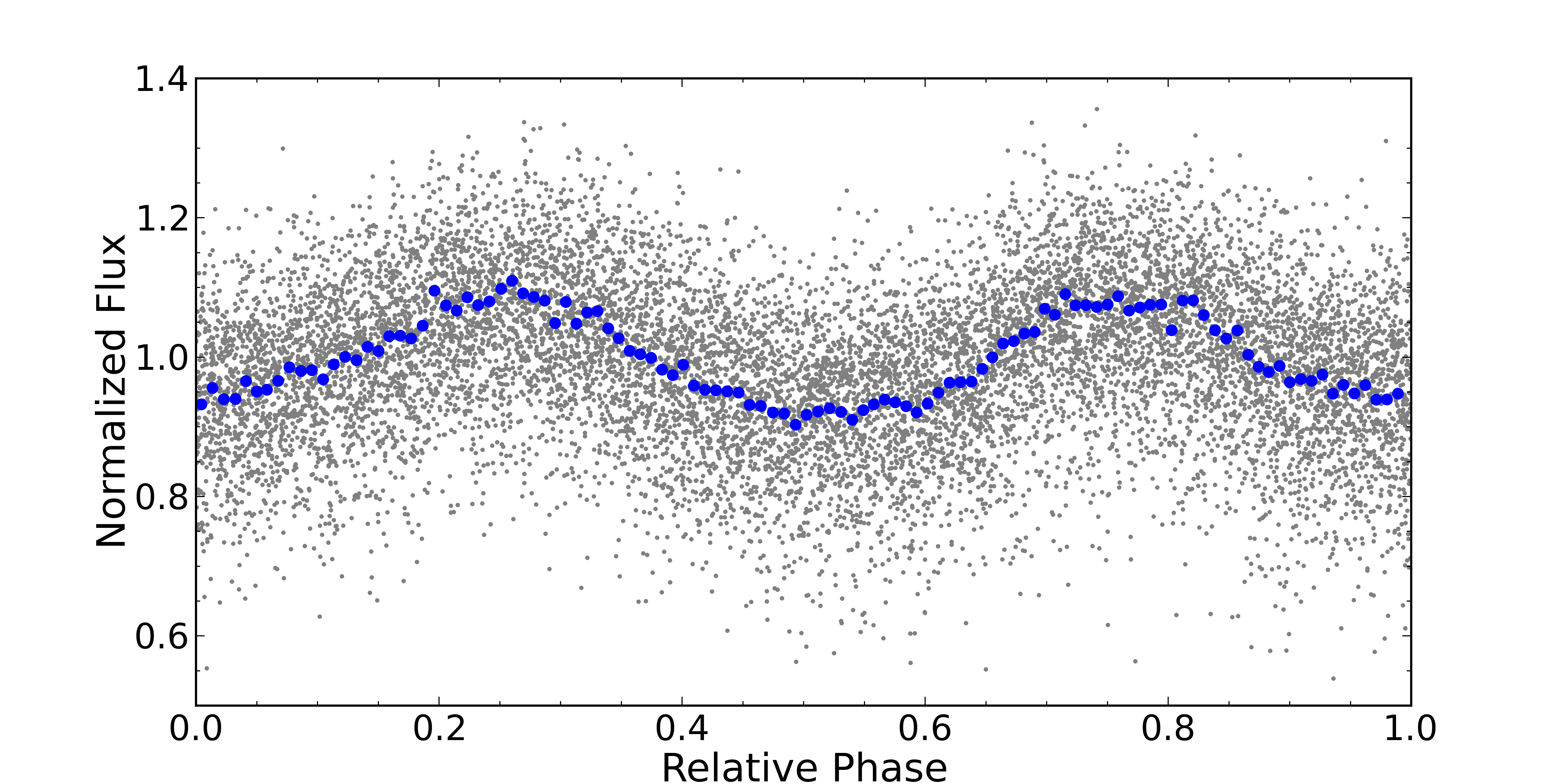}
\includegraphics[width=1.0\columnwidth]{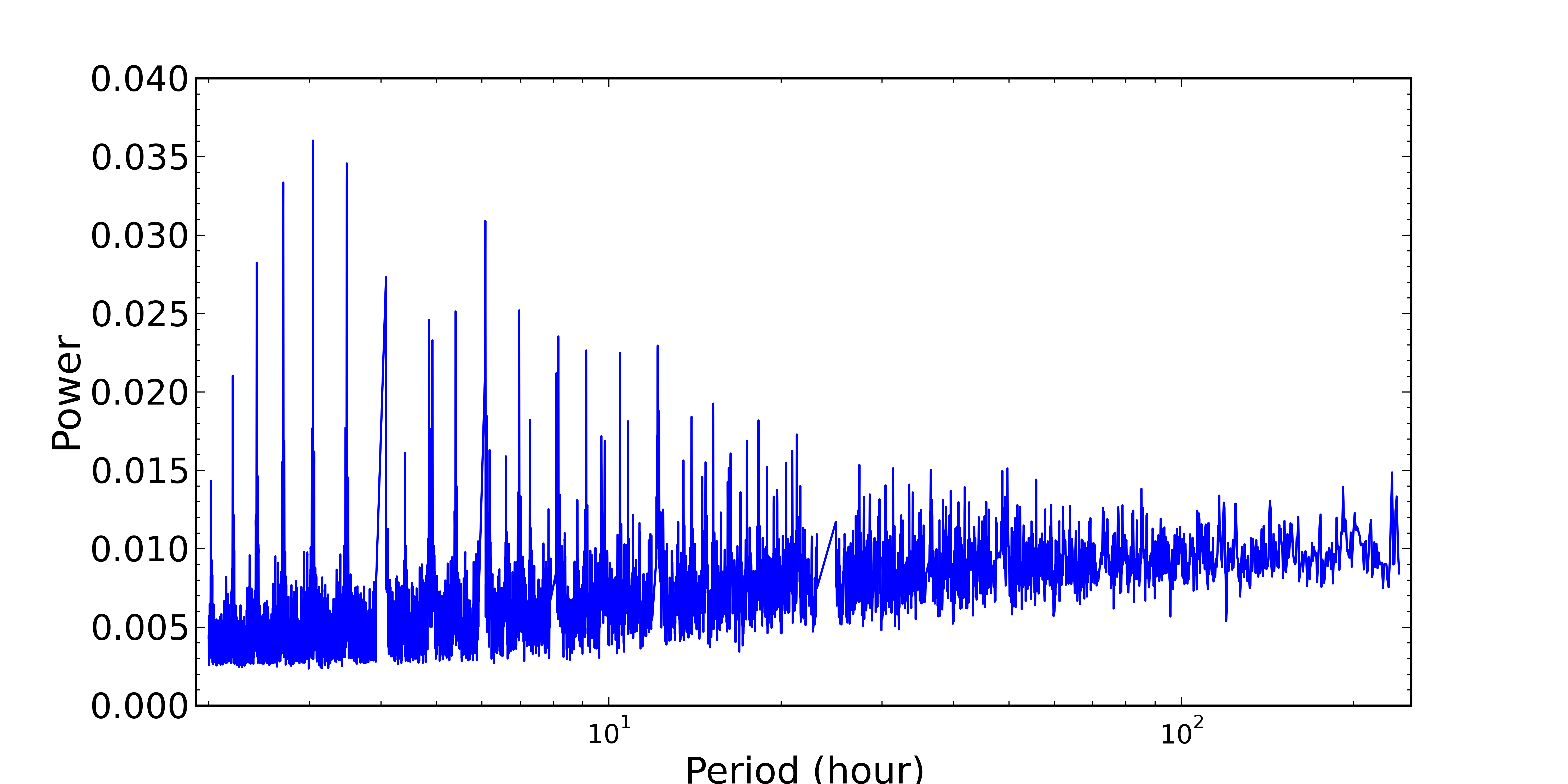}
\caption{The Evryscope discovery light curve of EVR-CB-004 folded on its period of 6.0846 hours is shown on the top panel. Grey points = 2 minute cadence, blue points = binned in phase. The bottom panel shows the BLS power spectrum with the highest peak at the 3.0423 hour detection (an alias of half of the actual period).}
\label{fig:EVR004_fig}
\end{figure}

We use a panel-detection plot that filters the light curves, identifies prominent systematics, searches a range of periods, and phase folds the best detections from several algorithms for visual inspection. It includes several matched filters to identify candidate hot subdwarfs for variability and is described in detail in \citep{2019PASP..131h4201R}. EVR-CB-004 was discovered using Box Least Squares (BLS; \citealt{Kovacs:2002gn, 2014A&A...561A.138O}) and Lomb-Scargle (LS) \citep{1982Ap&SS..263..835S} with the same settings, pre-filtering, and daily-alias masking described in \citet{2019PASP..131h4201R}. The discovery tools and settings were tested extensively to maximize recovery of the fast transits and eclipses characteristic of hot subdwarfs and white dwarfs. As part of our testing, we also recovered CD-30 11223 \citep{2041-8205-759-1-L25}, the only known fast-period hot subdwarf $+$ WD binary in our field of view and magnitude range, and discovered the compact evolving WD binary EVR-CB-001 \citep{2019ApJ...883...51R}. The BLS and LS power spectrum peaks correspond to 3.0423 hour and 3.04219 hour periods, respectively. Both detections found a period alias of half the actual period, and the candidate was originally thought to be a hot subdwarf reflection effect binary. Further analysis (see \S~\ref{section_analysis_lc}) showed the candidate to be a 6.08 hr compact binary exhibiting strong (16\%) modulations. Figure \ref{fig:EVR004_fig} presents both the BLS power spectrum and phase-folded light curve.

\subsection{SOAR/Goodman Photometry}\label{section_soar_photometry}

In order to obtain a higher signal-to-noise (S/N) light curve for modeling, we observed EVR-CB-004 on April 9, 2019 on the 4.1-m SOAR 4.1 m telescope at Cerro Pachon, Chile, with the Goodman spectrograph \citep{2004SPIE.5492..331C} in imaging mode. We used the blue camera with Bessel-V blocking filter, and took 515 images with 20 second exposure times. The image region of interest (ROI) was reduced to 1200 x 1200 pixels with 1x1 binning, which gave a 69\% duty cycle. For calibrations, we took 10 dome flats using 25\% lamp power and 10s integrations, 10 darks also with 10s integrations, and 10 bias images.

The SOAR images were processed with a custom aperture photometry pipeline written in Python. The images were dark and bias-subtracted and flat-field-corrected using the master calibration frames. Six reference stars of similar magnitude are selected and aperture photometry is performed using a range of aperture sizes. The background is estimated using the same size aperture for dark regions near each reference star. For full details of our SOAR photometry code, we refer the reader to \cite{2019ApJ...883...51R}. The resulting SOAR light curve is used to model EVR-CB-004 and check for eclipses and is shown later in the manuscript, in \S~\ref{section_analysis_lc}.

Since the TESS light curve is available for EVR-CB-004 (see the following section), the SOAR light curve provides an independent measurement in a much bluer band and is used as one of our two primary modeling solutions. The final solutions are consistent regardless of filter or instrument (see \S~\ref{section_analysis_lc}). The SOAR light curve was also used to rule out the shorter time scale eclipses (the TESS cadence is 2 min, while the SOAR cadence is 20 seconds, and expected eclipses would last $\approx$ 10 minutes).

We demonstrate later in the manuscript that EVR-CB-004 shows a small amplitude ($\approx.3\%$ in SOAR and $\approx.4\%$ in TESS) sinusoidal variation in the light curve, distinct from the main binary variability. This small amplitude variability is quite unexpected, and we needed to make sure it was not instrumental. The SOAR light curve is used to confirm this signal and measure it in a different band-pass to check for a wavelength dependent amplitude (see \S~\ref{section_low_amp_var}).

\subsection{TESS Photometry}\label{sec:tess}

EVR-CB-004 (TIC 1973623) was observed by TESS in Sector 8, from February 2-27, 2019, using Camera \#2. Photometry was obtained in the 120-second cadence mode and consists of 13,206 individual measurements spanning 24.5 days, including a short interruption near the middle of the sequence to allow for the data to be downlinked. We use for our analysis the presearch data conditioning (PDC) light curve extraction \citep{Smith2012,Stumpe2014} provided by the TESS Science Processing Operations Center \citep{Jenkins2016}. These data are made publicly available through the Mikulski Archive for Space Telescopes. A LS periodogram analysis shows a clear detection of the 6.08 d binary signal and its harmonics. We find no other statistically significant peaks out to the Nyquist frequency (360 d$^{-1}$) and limit additional variability to amplitudes $<$550 ppm. We used the TESS light curve for our light curve analysis solution with a red band-pass, independent from the SOAR light curve.

The coarse TESS pixel scale is prone to blending from nearby stars, potentially contaminating the signal from the target. The very fine SOAR pixels (.15" per pixel) easily resolve nearby stars in the field, and the SOAR image revealed three nearby stars that were potential contaminants in the TESS pixel. Simple tests (see \S~\ref{section_discussion}) showed these to be constant, much lower in flux than the target, and to not affect the light curve or solution to EVR-CB-004.

\subsection{PROMPT Photometry}

We observed EVR-CB-004 with the PROMPT MO1 46cm telescope \citep{2005NCimC..28..767R} located at Meckering Australia, in Johnson R band. The PROMPT photometric observations provided an intermediate filter to the SOAR and TESS data, and verified the light curve solution in \S~\ref{section_analysis_lc}. The observations were taken on March 30, 2019, continuously over the period with 120 s exposure times. We also obtained bias, flat, and dark calibration images. The images were processed with a custom pipeline that uses standard calibration and aperture photometry, using 5 nearby reference stars of similar magnitude to correct for airmass and observing conditions. For a detailed description of the pipeline, we refer the reader to \cite{2019PASP..131h4201R}.

\subsection{SMARTS 1.5-m/CHIRON Spectroscopy}

We observed EVR-CB-004 with the SMARTS 1.5 m telescope and CHIRON, a fiber-fed cross-dispersed echelle spectrometer \citep{2013PASP..125.1336T}. Six spectra were obtained in image fiber mode (R $\sim$ 28,000) between March and July 2019 and  covered the wavelength range 4400-8800 \AA. We used integration times of 1200 s to obtain just enough S/N for radial velocity (RV) measurements; longer integrations would have resulted in too much phase-smearing. All raw spectra were reduced and wavelength-calibrated by the official CHIRON pipeline, housed at Georgia State University and managed by the SMARTS Consortium\footnote{http://www.astro.yale.edu/smarts/}. In addition to  H$\alpha$ and H$\beta$, which span multiple orders, the spectra show four He\,{\sc i} lines, including 6678 \AA, 5876 \AA, 5016 \AA, and 4922 \AA, and two He\,{\sc ii} lines 4686 \AA\ and 5412 \AA. All of these features are synced in phase, with no signs of absorption due to a companion, and we conclude they emanate from a single star.

\subsection{SOAR/Goodman Spectroscopy}

\subsubsection{Low-Resolution (for Atmospheric Modeling)}

We obtained low-resolution spectra for atmospheric modeling on February 9, 2019 with the Goodman spectrograph using the 600 mm$^{-1}$ grating blue preset mode, 2x2 binning, and the 1" slit. This configuration provided a wavelength coverage of 3500-6000 \AA\ with spectral resolution of 4.3 \AA\ (R$\sim$1150 at 5000 \AA). We took four 360 s spectra of both the target and the spectrophotometric standard star BPM 16274. For calibrations, we obtained 3 x 60 s FeAr lamps, 10 internal quartz flats using 50\% quartz power and 30 s integrations, and 10 bias frames.

We processed the spectra with a custom pipeline written in Python, described in \cite{2019ApJ...883...51R}. Each of the processed spectra was then rest-wavelength calibrated using a Gaussian fit to the H$\beta$ through H11 absorption features, as well as several prominent He absorption features. The resulting spectra were median-combined to form a final spectrum for atmospheric modeling. As shown in Figure \ref{fig:EVR004_specfit_low}, we detect strong H Balmer lines, from H$\beta$ through H13, and several He lines. We find no evidence of absorption features due the companion star; at this resolution EVR-CB-004 is a single-lined binary.

\subsubsection{Medium-Resolution (for Radial Velocity)} \label{section_soar_rv_spec}

To measure the RV of EVR-CB-004, we also obtained medium-resolution spectra on March 5, 2019 with the Goodman spectrograph using the 2100 mm$^{-1}$ grating in custom mode, 1x2 binning, and the 0.46" slit. This configuration provided a wavelength coverage of 3700-4400 \AA\ with spectral resolution of 0.34 \AA\ (R$\sim$11930 at 4000 \AA). We took 32 x 360 s spectra of the target and 3 x 60 s FeAr lamps after every fourth spectrum. We observed uninterrupted to cover the half of the period from minimum to maximum. For calibrations, we obtained 10 internal quartz flats using 80\% quartz power and 60 s integrations, and 10 bias frames.

We processed the spectra with a custom pipeline written in Python, described in \cite{2019ApJ...883...51R}. The groups of 4 processed spectra were median-combined to form a final spectrum used to determine the RV. As shown in Figure \ref{fig:EVR004_specfit_medium}, We detect strong H Balmer lines, from H$\gamma$ through H10, and several He lines. In this resolution mode we also find CaH and CaK lines that originate from a different source than all other features. We discuss the origin of the Ca lines in \S~\ref{section_discussion}.

\begin{table*}[t]
\caption{Overview of Observations for EVR-CB-004}
\centering
\begin{tabular}{l l l l l}
\hline
Telescope & Date & Filter/Resolution & Epochs & Exposure\\
\hline
Photometry & & & &\\
Evryscope & Jan 2017 - Jun 2017 & Sloan \textit{g} & 4,812 & 2 min\\
SOAR/Goodman & April 9, 2019 & Bessel-V & 515 & 20 s\\
TESS & Feb 2-27, 2019 & 600-1000\textit{nm} & 13,206 & 2 min\\
PROMPT & March 30, 2019 & Johnson-R & 180 & 2 min\\
\hline
Spectroscopy & & & &\\
SMARTS 1.5-m/CHIRON & Mar 2019 Jul 2019 & 28,000 & 6 & 1200 s\\
SOAR/Goodman & Feb 9, 2019 & 1150 & 4 & 360 s\\
SOAR/Goodman & March 5, 2019 & 11930 & 32 & 360 s\\

\hline
\end{tabular}
\label{observations_table}
\end{table*}

Table \ref{observations_table} presents a brief overview of all of the photometric and spectroscopic data used in our analysis of EVR-CB-004.


\section{ORBITAL AND ATMOSPHERIC PARAMETERS} \label{section_analysis_spectra}

To measure radial velocities (RVs), we first inspected the SOAR spectra (see \S~\ref{section_soar_rv_spec}) and selected prominent absorption features with the highest signal to noise, found to be H$\gamma$-H10. These features (3750\AA, 3835\AA, 3889\AA, 3970\AA, 4102\AA, 4340\AA) are then used for fitting, by clipping small regions encompassing each absorption line and measuring the central value using a Gaussian fit. We measure the shift, calculate the velocity, and use the standard deviation in the velocities of the 6 absorption features to determine the uncertainty. The resulting velocities were converted to heliocentric velocities using PyAstronomy's \textit{baryCorr} function.

\begin{figure}[ht]
\includegraphics[width=1.0\columnwidth]{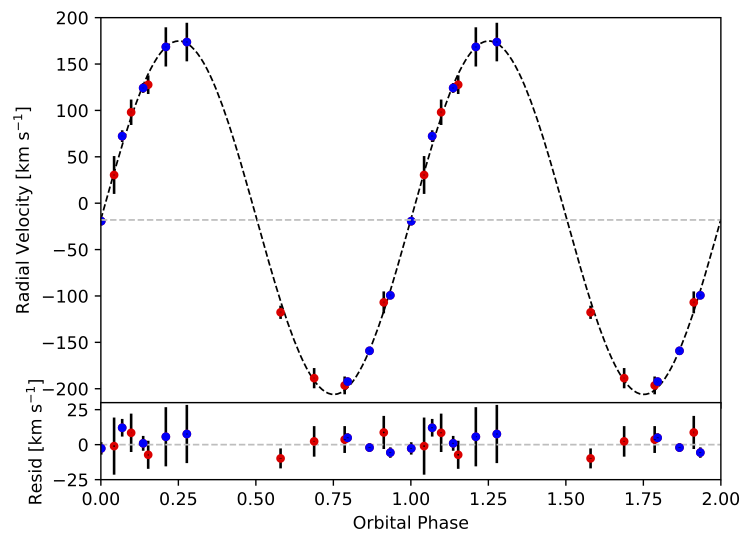}
\caption{{\em Top panel:} Phase-folded, heliocentric radial velocity measurements from SMARTS 1.5-m/CHIRON (red) and SOAR/Goodman (blue), plotted twice for better visualization. The black dashed line denotes the best-fitting sine wave to the data. After correcting for slight phase smearing, we find a velocity semi-amplitude of K = 190.5 $\pm$ 2.8 km s$^{−1}$ and a systemic velocity of $\gamma$ = -18 $\pm$ 4 km s$^{−1}$. {\em Bottom panel:} Residuals after subtracting the best-fitting sine wave from the data.}
\label{fig:EVR004_RV_fig}
\end{figure}

The CHIRON spectra are processed in a similar way but using the absorption features falling in the CHIRON wavelength coverage. The CHIRON and Goodman measurements were combined together and phase-folded using the period determined from the light curve. With the period and phase fixed to values determined from the photometry, we fitted a sine wave to the radial velocity curve and find a semi--amplitude of K = 190.5 $\pm$ 2.8 km/s. Figure \ref{fig:EVR004_RV_fig} presents the radial velocity curve and best--fitting sine wave.

\begin{figure*}[ht]
\centering
\includegraphics[width=0.68\columnwidth]{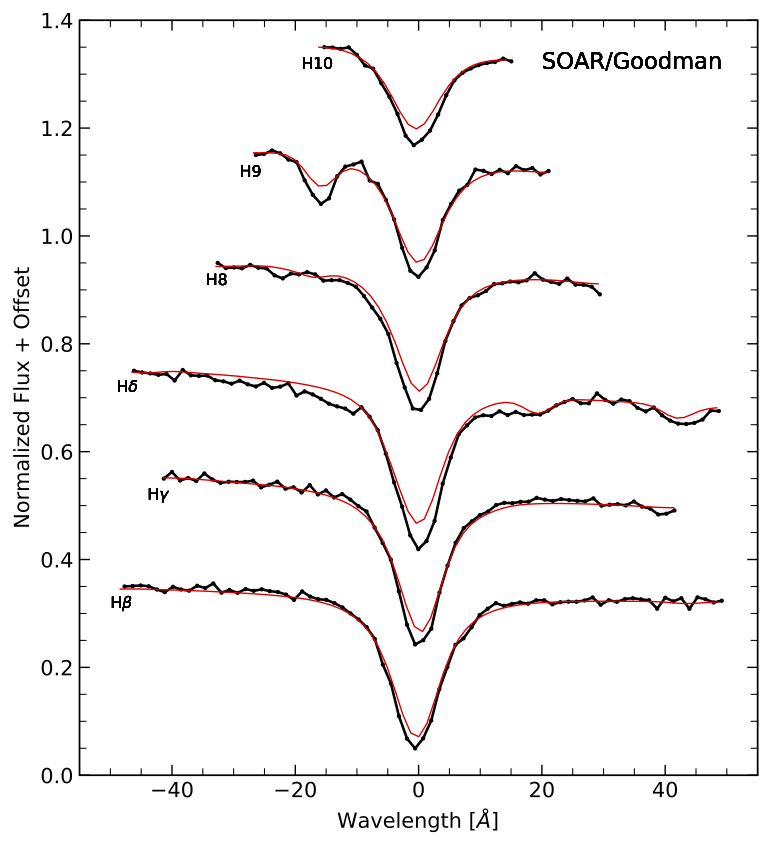}
\includegraphics[width=0.68\columnwidth]{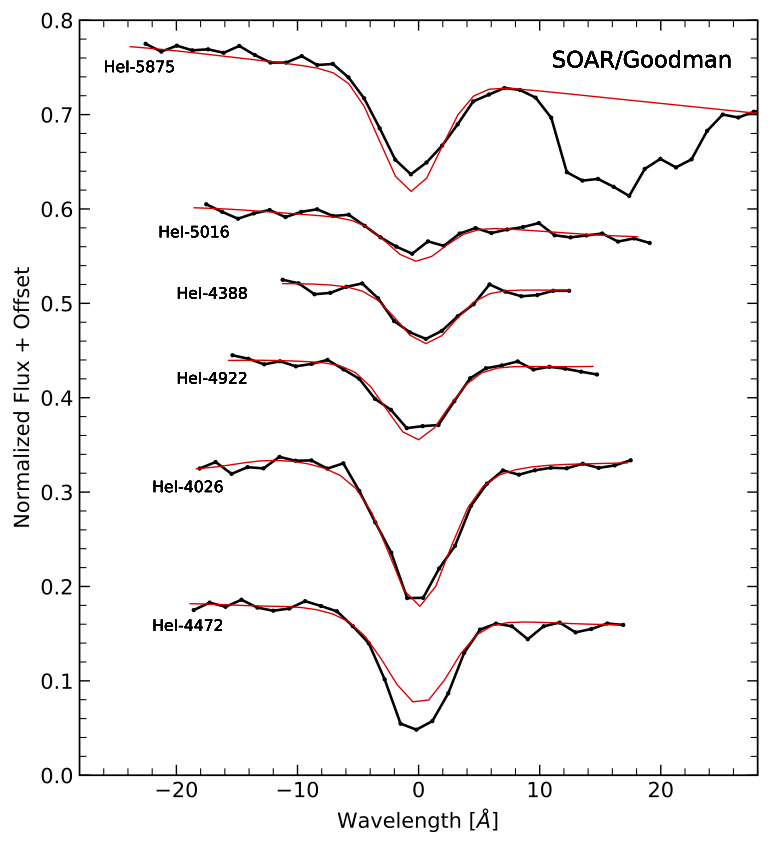}
\includegraphics[width=0.68\columnwidth]{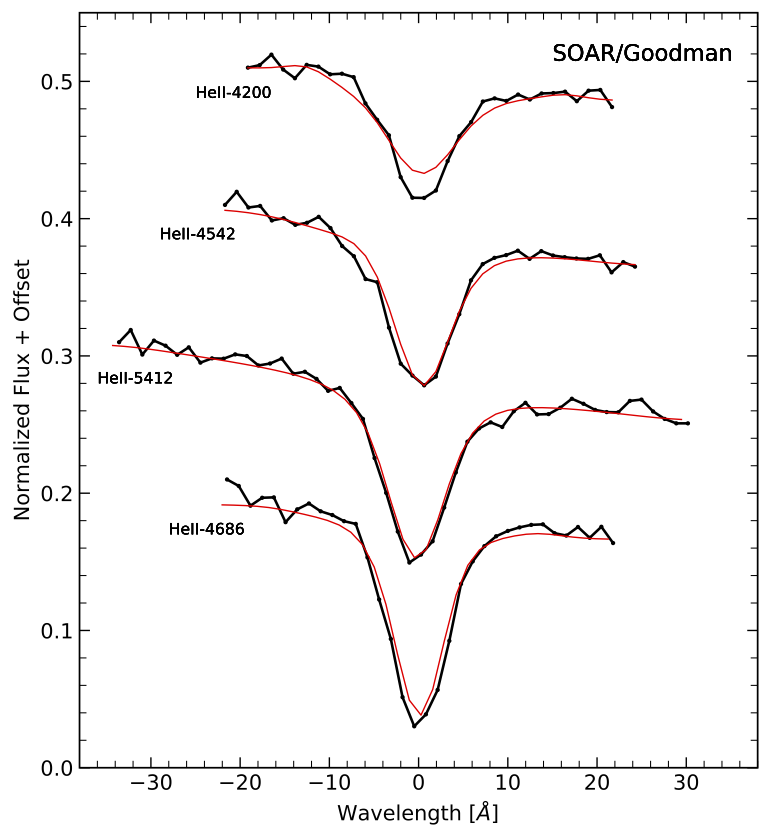}
\caption{Normalized and stacked low-resolution SOAR/Goodman spectrum of EVR-CB-004 (black line) with best--fitting atmospheric model (red line). The panels highlight H Balmer (left), He \textsc{i} lines (middle), and He \textsc{ii} (right) absorption features.}
\label{fig:EVR004_specfit_low}
\end{figure*}

\begin{figure}[ht]
\centering
\includegraphics[width=1\columnwidth]{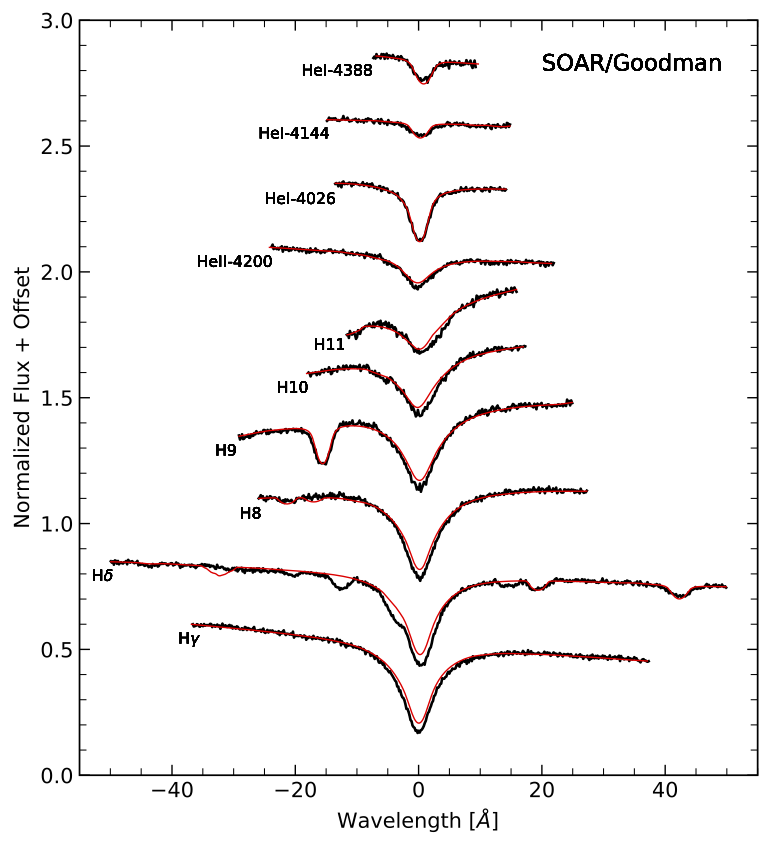}

\caption{Normalized and stacked medium-resolution SOAR/Goodman spectrum of EVR-CB-004 (black line) with best--fitting atmospheric model (red line).}
\label{fig:EVR004_specfit_medium}
\end{figure}

Because the H Balmer lines span multiple orders in the high-resolution CHIRON spectra, critical features including the continuum and absorption lines are segmented, making them insufficient to determine reliable atmospheric parameters (effective temperature $T_{\text{eff}}$, surface gravity $\log{(g)}$, and helium abundance $\log{n(\text{He})}=\log{N(\text{He})/N(H)}$). They were also not suitable to determine the projected rotational velocity $v_{\text{rot}}\sin{i}$ due to phase smearing caused by the necessarily long exposure times. Therefore, we Doppler-corrected all SOAR RV spectra to the same rest frame and stacked them to create a master medium-resolution spectrum as done for the low-resolution SOAR data. We then used both SOAR resolutions for our spectroscopic analysis.

To determine the atmospheric parameters, we fitted the observed H and He line profiles simultaneously (see Fig. \ref{fig:EVR004_specfit_low} \& \ref{fig:EVR004_specfit_medium}). The rotational velocity $v_{\text{rot}}\sin{i}$ was determined from the average medium-resolution spectrum only. The H Balmer lines closest to the Balmer jump were of special interest to us since they are most sensitive to $\log(g)$ and $T_{\mathrm{eff}}$. We calculated a grid of non-local thermodynamic equilibrium (NLTE) model atmospheres with TLUSTY 205 and the spectral synthesis was realized with SYNSPEC 51 \citep{1988CoPhC..52..103H, 1995ApJ...439..875H, 2003ASPC..288...51H, 2017arXiv170601859H, 2017arXiv170601935H, 2017arXiv170601937H}. Radiative and hydrostatic equilibrium, plane-parallel geometry as well as chemical homogeneity were assumed. The temperature and density stratification in the hydrogen and helium line-forming regions were well constrained, once carbon, nitrogen, and oxygen were included as absorbers (see also \citealt{2018A&A...620A..36S} for details). These models are also less computationally demanding than more complex models, such as those including iron and nickel, where the added features do not materially change the outcome but are very computationally intensive. Making use of the detailed model atoms listed in Table \ref{TLUSTY/SYNSPEC Ionization stages used}, the following ionization stages with mean metallicities for hot subdwarf B stars from \citet{2013MNRAS.434.1920N} were synthesized: H\,{\sc i}, He\,{\sc i/ii}, C\,{\sc ii/iii/iv}, N\,{\sc ii/iii/iv/v}, and O\,{\sc ii/iii/iv}. For each element, the ground state of the next higher ionization stage was also included. Stark broadening tables for H\,{\sc i} according to \citet{2009ApJ...696.1755T}, for He\,{\sc i} according to \citet{1969PhDT........47S} and \citet{1974JQSRT..14.1025B}, and for He\,{\sc ii} according to \citet{1989A&AS...78...51S} were used.

The selective fitting routine used is based on the FITSB2 spectral analysis program \citep{2004ASPC..318..402N}, the ``Spectrum Plotting and Analysis Suite'' SPAS \citep{2009PhDT.......273H}, and the $\chi^2$-based fitting procedure described by \citet{1999A&A...350..101N}. Cubic spline interpolation was used to interpolate between different synthetic spectra and the actual fit to the preselected hydrogen and helium lines in the observed spectrum was performed via the downhill \textit{simplex} algorithm from \citet{Nelder_and_Mead_1965}. The continuum was set at the edges of the preselected lines and the synthetic spectrum was folded with the instrumental profile.

From our NLTE quantitative spectral analysis, we were able to fit the He \textsc{i} and He \textsc{ii} lines consistently, indicating that $T_{\text{eff}}$ is well constrained. The Balmer line wings could be matched, but there is no way to fit the cores simultaneously (see Fig.~\ref{fig:EVR004_specfit_low} \& \ref{fig:EVR004_specfit_medium}). This is most likely due to shortcomings of the model atmospheres which do not include metal line blanketing beyond carbon, nitrogen, and oxygen. However, the derived surface gravity is reliable, since the Balmer line wings are matched reasonably well.

We found $v_{\text{rot}}\sin{i}=116.5\pm8.1$\,km\,s$^{-1}$, $T_{\mathrm{eff}}=41000\pm200$\,K, $\log{(g)}=4.55\pm0.03$, and $\log{n(\text{He})}=-0.84\pm0.02$ from the medium-resolution and $T_{\mathrm{eff}}=41500\pm1100$\,K, $\log{(g)}=4.60\pm0.12$, and $\log{n(\text{He})}=-0.90\pm0.09$ from the low-resolution SOAR data. We note that here and throughout the rest of the manuscript we follow the convention in expressing $\log{(g)}$ values as the unit-less surface gravity, understood shorthand for $\log{(g/cm s^{-2})}$ with g expressed in units of $cm s^{-2}$. In the low-resolution SOAR data, we fixed $v_{\text{rot}}\sin{i}$ to the value derived from the medium-resolution spectrum. The 1$\sigma$ statistical errors given above were derived using a simple bootstrapping method, whereby the data themselves were randomly resampled with replacement a large number of times and a parameter fit for each of the iterations was performed. Finally, the $1\sigma$ standard error for each parameter was derived from the standard deviation of the respective parameter bootstrap distribution.

Due to the near perfect agreement between the low and medium-resolution results, we took the weighted averages of each of the atmospheric parameters derived and consider them as the final results of the atmospheric modeling. Table \ref{solutions_table} lists them: $T_{\mathrm{eff}}=41016\pm197$\,K, $\log{(g)}=4.553\pm0.030$, and $\log{n(\text{He})}=-0.843\pm0.020$ (1$\sigma$ statistical errors only). The error budget on the atmospheric parameters is not dominated by statistical, but rather by systematic uncertainties, which are always difficult to estimate in spectroscopy. We decided to use $\Delta T_{\text{eff}}/T_{\text{eff}}=$\,3\,\%, $\Delta\log{(g)}=0.10$, and $\Delta\log{n(\text{He})}=0.13$, which is rather conservative. 

From our observed spectra and derived atmospheric parameters alone the primary star in EVR-CB-004 is an O-type. To more precisely determine the stellar classification, we consider the luminosity of the primary in EVR-CB-004 which we calculate to be $\log{(L_{\text{Gaia}}/L_{\odot})} = 2.97 \pm 0.11$, using the \textit{Gaia} parallax and Magnitude. See Table \ref{solutions_table_spma} later in the manuscript listing the fundamental parameters of the primary of EVR-CB-004. We would expect a main-sequence O-star to have a much higher luminosity of at least $\log{(L_{\text{Gaia}}/L_{\odot})}$ = 5 to 6. The primary of EVR-CB-004 is sub-luminous for its high temperature. From the spectra, temperature, surface gravity, and luminosity, we classify the primary in EVR-CB-004 as an sdO hot subdwarf.

Later in the manuscript, in \S~\ref{section_system_solution}, the mass of the primary, the light curve features, the period, and the separation of the system are shown to be consistent with a hot subdwarf. Since the hot subdwarf sdO label includes stars with different evolutionary histories, we discuss in \S~\ref{section_discussion} how the low-surface gravity of the primary in EVR-CB-004 is helpful in understanding its current evolutionary status and history.

\begin{table}
\caption{Ionization stages for which detailed model atoms were used in the model atmosphere calculations for TLUSTY/SYNSPEC. The number of levels (L) and super-levels (SL) is listed. For each element the ground state of the next higher ionization stage was also included, but is not listed here.}\label{TLUSTY/SYNSPEC Ionization stages used}
\centering
\begin{tabular}{ccc|ccc}
\hline\hline
Ion & L & SL & Ion & L & SL\\
\hline
H\,{\sc i} & 16 & 1 & N\,{\sc iii} & 25 & 7\\
He\,{\sc i} & 24 & 0 & N\,{\sc iv} & 34 & 14\\
He\,{\sc ii} & 20 & 0 & N\,{\sc v} & 10 & 6\\
C\,{\sc ii} & 17 & 5 & O\,{\sc ii} & 36 & 12\\
C\,{\sc iii} & 34 & 12 & O\,{\sc iii} & 28 & 13\\
C\,{\sc iv} & 21 & 4 & O\,{\sc iv} & 31 & 8\\
N\,{\sc ii} & 32 & 10 & & &\\
\hline
\end{tabular}
\end{table}\noindent

\begin{figure*}[ht]
\centering
\includegraphics[width=0.69\columnwidth]{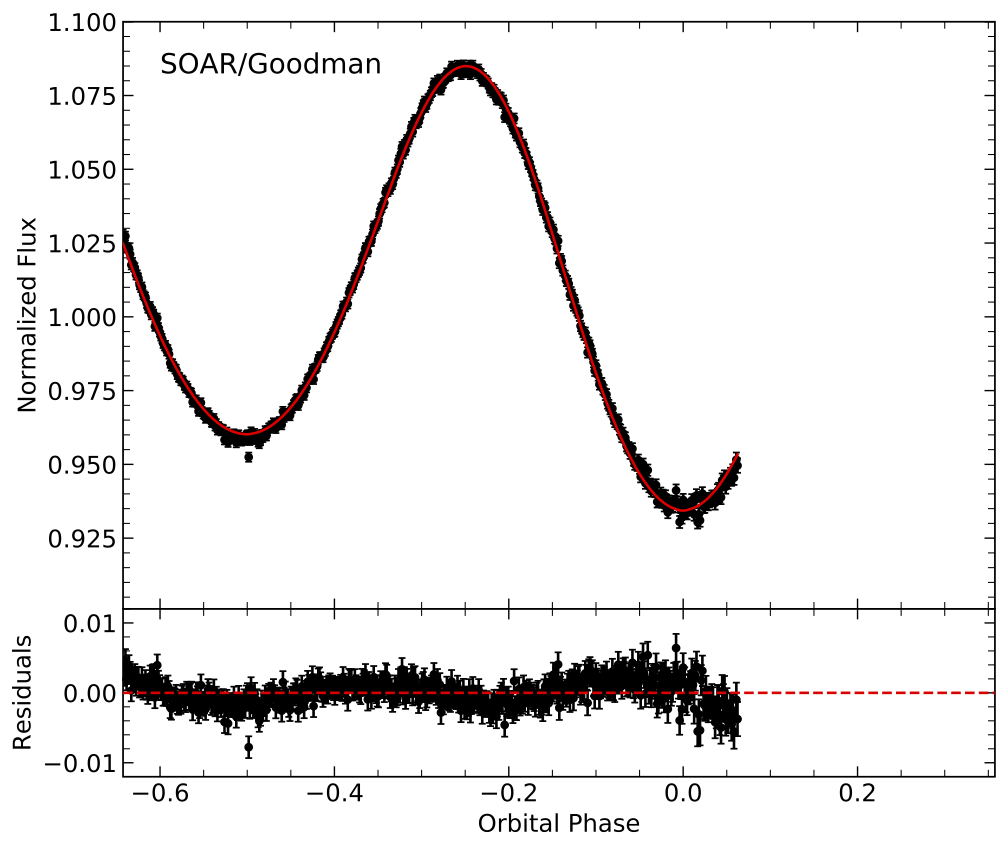}
\includegraphics[width=0.69\columnwidth]{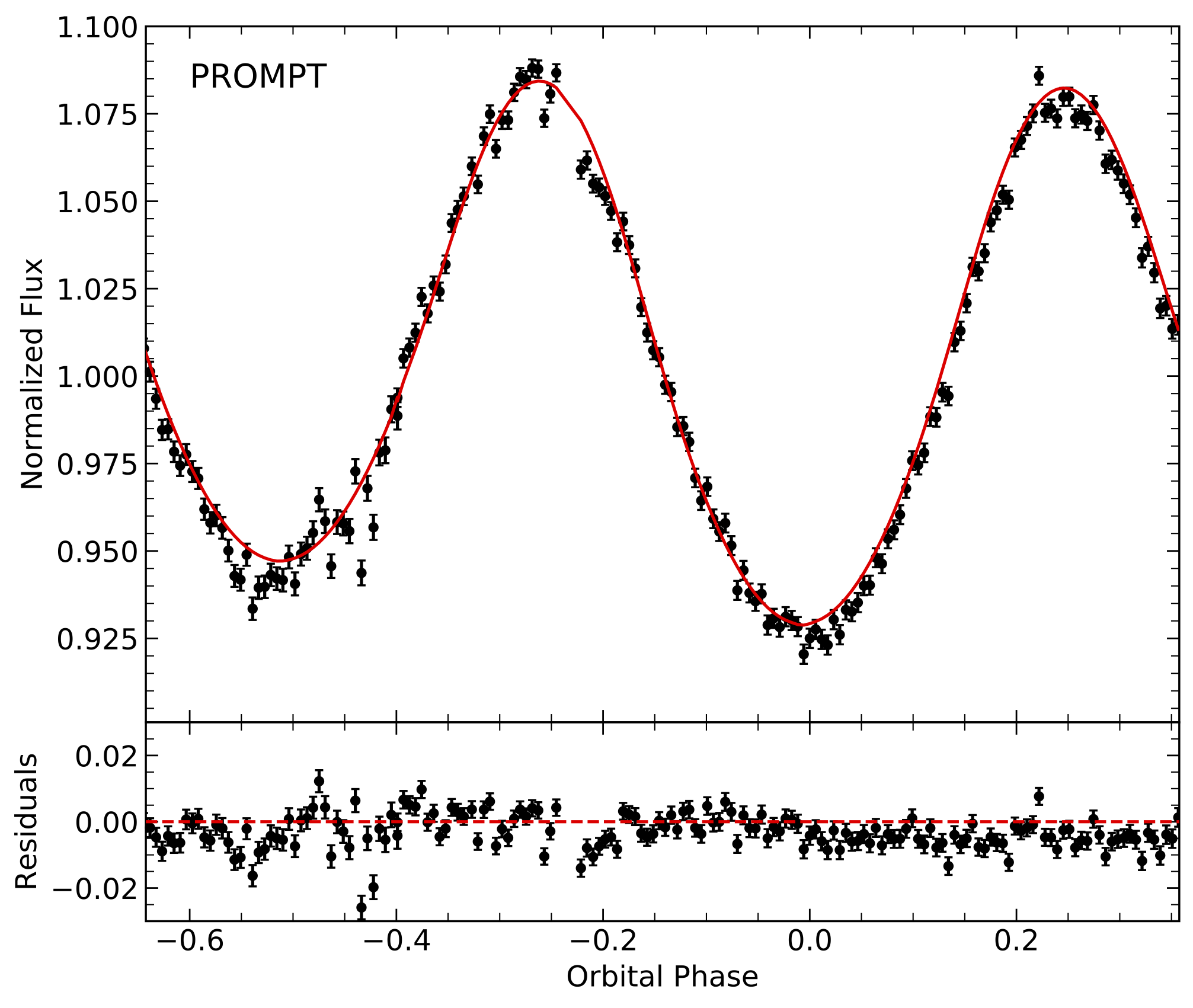}
\includegraphics[width=0.69\columnwidth]{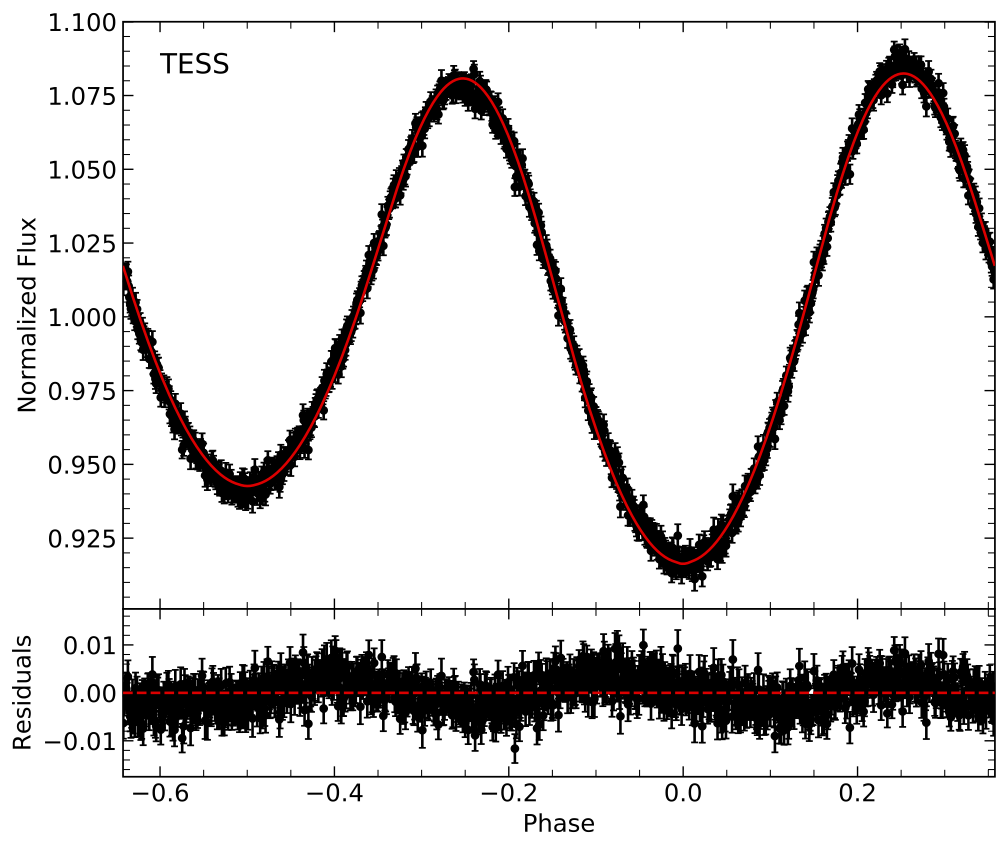}
\caption{The SOAR/Goodman (left; V filter), PROMPT (middle; R filter), and TESS (right; $\sim$I filter) light curves with the best-fitting model determined from {\sc lcurve}. The best--fitting model was determined from simultaneous fits to all three light curves. The PROMPT and SOAR data were taken continuously, while the TESS light curve shown was produced by phase--folding and binning the full 27-d light curve. The residuals show a coherent signal at 1/3 the orbital period, which is discussed in Section \ref{section_low_amp_var}.}
\label{fig:EVR004_lcfit}
\end{figure*}


\begin{table}
 \centering
 \caption{Overview of the fixed parameters for the LCURVE fit}
  \begin{tabular}{lrrrrr}
  \hline\hline
  Parameter &   TESS & PROMPT  & SOAR  \\
            &  I-band  & R-band & V-band \\
  \hline
  Beaming Factor (F)      &    1.24   &  1.30   &  1.35    \\
  gravity darkening $\beta$   &  0.26  &  0.26  &  0.27   \\
  limb darkening $a_\mathrm{1}$    &    1.34    &   1.39   &  1.38   \\   
  limb darkening $a_\mathrm{2}$   &     -2.25   &  -2.23  & -2.06     \\
  limb darkening $a_\mathrm{3}$   &    2.03   &  1.97   &  1.79    \\
  limb darkening $a_\mathrm{4}$   &    -0.69    &  -0.66  &  -0.595      \\
     \hline
\end{tabular}
\label{tab:fitparam}
\end{table}

\section{LIGHT CURVE ANALYSIS} \label{section_analysis_lc}

Since only spectral features from the sdO primary star are detected, we must rely on light curve modeling to compute the mass ratio $q$ and constrain the system's parameters. We use the modeling code {\sc lcurve} \citep{2010MNRAS.402.1824C} to analyze the TESS I-band, SOAR V-band and PROMPT R-band light curves. We assume that the orbit is circular. The flux that each point on the grid emits is calculated by assuming a blackbody of a certain temperature at the bandpass wavelength, corrected for limb darkening, gravity darkening, Doppler beaming and the reflection effect.

The strong (16\%) modulations in the light curve are due to the ellipsoidal deformation of the primary from the unseen, more massive companion. A subtle asymmetry (a sub 1\% difference in the height of alternating peaks) is observed, indicative of Doppler boosting with the higher peak corresponding to the orbital position where the sdO is moving toward us most quickly. The difference in minima is due to gravitational darkening of the deformed sdO, with the lower minimum corresponding to the orbital position where the sdO is farthest from us.

The light curve of EVR-CB-004 is dominated by ellipsoidal modulations due to tidal distortion of the sdO star. Ellipsoidal modulations are sensitive to the mass ratio, the size of the distorted star relative to the orbital separation and the limb and gravity darkening \citep{morris1985}. For the invisible companion we assume a lower limit to the radius, using the mass-radius relation for fully degenerate WDs by Eggleton (quoted from \citealt{verbunt88}). The general 4-parameter limb darkening prescription and the passband specific gravity darkening prescription were used following \citet{cla04,blo11} and as tabulated in \citet{claret11}. The values used for the beaming, limb darkening and gravity darkening are shown in Table.\,\ref{tab:fitparam}.  Additionally, we added a constant third light component to the TESS light curve to account for the contributions from the close-by stars (see \S~\ref{sec:tess}) and a first order polynomial to the SOAR and PROMPT lightcurve to account for an airmass effect.

Using the results for surface gravity ($\log{g}$), effective temperature ($T_{\mathrm{eff}}$), combined with the orbital period ($P$) and radial velocity ($K_{\rm sdO}$), we determine the inclination angle ($i$), the mass ratio ($q$), the secondary temperature $T_{\rm WD}$, as well as the scaled radii and velocity scale (($K_{\rm sdO}+K_{\rm WD})/\sin{i}$). The subscript sdO is used for the sdO star which dominates the light ($K_{\rm sdO},M_{\rm sdO},R_{\rm sdO}$), and the subscript WD is used for the invisible companion ($K_{\rm WD},M_{\rm WD},R_{\rm WD}$).

Using this model we were not able to find a consistent solution with flat residual. In each light curve we find a coherent signal at 1/3 the orbital period with a low amplitude of $\approx$0.5\,\%. We obtained a reduced $\chi^2 \approx 1.5$. Even allowing the limb, gravity darkening coefficients or the beaming factor to float free (and to iterate towards implausible values), the residuals remain in the light curve fit. We discuss possible explanations for the residuals in \S~\ref{section_low_amp_var}.

We combine {\sc lcurve} with the MCMC implementation {\sc emcee} \citep{2013PASP..125..306F} to explore the parameter space, converge on a solution, and to determine the uncertainties. We used 256 chains and let them run for 2000 trials well beyond a stable solution was reached. The corner plot of the final solution is shown in Figure\,\ref{fig:mcmc_results} in the appendix. The final fits using the TESS, SOAR and PROMPT light curve are shown in Figure \ref{fig:EVR004_lcfit}. The ellipsoidal deformation dominates the photometric variation in the light curve, but Doppler boosting and gravity darkening effects are also present.\\


\section{RESULTS} \label{section_system_solution}

Although EVR-CB-004 is a single-lined binary, we can still constrain the masses and radii of the two stars by combining the results of the light curve modeling with results from the spectroscopic fitting. Parameters derived in this way by a simultaneous fit to the SOAR, PROMPT and TESS light curves are summarized in Table \ref{solutions_table}.

Our solution converges on a mass ratio of $q$ = $M_{\rm sdO}/M_{\rm WD}$ = 0.76$\pm$0.03, with individual masses of $M_{\rm sdO}$ = 0.52$\pm$0.04 M$_{\odot}$ and $M_{\rm WD}$ = 0.68$\pm$0.03 M$_{\odot}$ . We reiterate that the sdO star is the dominant source of light in the system, and the one showing ellipsoidal modulation. This object has a radius of $R_{\rm sdO} = 0.63\pm0.02\,R_{\odot}$, much larger than the canonical radius of most known sdO stars. We refer to this as inflated hereafter and note specifically that this is without any suggestion to internal structure or non-equilibrium mechanism. We find a Roche Lobe filling factor ($f = R_{\rm sdO}/R_L = 0.99\pm0.01$), where $R_L$ is the Roche Radius, close to $1$ and consistent with $1$ which shows that the sdO is close to filling its Roche Lobe and even consistent with filling its Roche Lobe entirely. The radius ($R_{\rm WD}$) of the unseen companion cannot be determined, due to the lack of eclipses. However, since it does not produce any detectable light in the system despite its higher mass, the companion is consistent with a WD. 

From the system parameters we find that the sdO should have a projected rotational velocity \vrot$=118\pm5$\,\kms\, to be synchronized to the orbit. The measured \vrot$=116.5\pm8.1$\,\kms\, is consistent with the predicted value and therefore we conclude that the sdO exhibits synchronous rotation as expected in a compact post-CE binary.

\begin{table*}[ht]
\caption{EVR-CB-004 Parameters. $\dagger$: 1$\sigma$ statistical errors only.}
\centering
\begin{tabular}{l l l l}
\hline
Description & Identifier & Units & Value\\
\hline
Evryscope ID & EVR-CB-004 &    &   \\
GAIADR2 ID & GAIADR25642627428172190000 &    &   \\
Right ascension & RA &  [degrees]  & 133.30233\\
Declination & Dec &  [degrees]  & -28.76838\\
Magnitude & $m_g$ & [mag] & 13.127 $\pm.002$\\
\hline
Hot Subdwarf Atmospheric Parameters &    &    &   \\
\hline
Effective temperature & $T_{\text{eff}}$ & [K] & $\mathrm{41\,016\pm197}^\dagger$\\
Surface gravity & $\log{(g)}$ &  & $\mathrm{4.553\pm0.030}^\dagger$\\
Helium abundance & $\log{n(\text{He})}=\log{N(\text{He})/N(H)}$ &    & $\mathrm{-0.843\pm0.020}^\dagger$\\
Projected rotational velocity & $v_{\text{rot}}\sin{i}$ & [km\,s$^{-1}$] & $116.5\pm8.1^\dagger$\\
\hline
Orbital Parameters\\
\hline
Orbital Period & P & [hours] & 6.0842 $\pm.0001$\\
RV semi-amplitude & H & [km s$^{-1}$] & 190.5 $\pm$ 2.8\\
System velocity & $\gamma$ & [km s$^{-1}$] & 18 $\pm$ 4\\
\hline
Solved Parameters\\
\hline
mass ratio & $q = \frac{M_{\rm sdO}}{M_{\rm WD}}$ &  &  0.76$\pm$0.03 \\
Hot subdwarf mass & $M_{\rm sdO}$ & [$M_{\odot}$] & 0.52$\pm$0.04 \\
Hot subdwarf radius & $R_{\rm sdO}$ & [$R_{\odot}$] & 0.63$\pm$0.02 \\
White dwarf mass & $M_{\rm wd}$ & [$M_{\odot}$] & 0.68$\pm$0.03 \\
Orbital Inclination & \textit{i} & [$^\circ$] & 69.9$\pm$1.0\\
Separation & \textit{a} & [$R_{\odot}$] & 1.79$\pm$0.03 \\
\hline
\multicolumn{4}{l}{We note that here and throughout the rest of the manuscript we follow the convention in}\\
\multicolumn{4}{l}{  expressing $\log{(g)}$ values as the unit-less surface gravity, understood}\\
\multicolumn{4}{l}{  shorthand for $\log{(g/cm s^{-2})}$ with g expressed in units of $cm s^{-2}$.}\\

\label{solutions_table}
\end{tabular}
\end{table*}


\section{DISCUSSION} \label{section_discussion}

\subsection{Independent mass estimate of the hot subdwarf - The Spectrophotometric Approach}\label{The Spectrophotometric Approach}

We measured the mass and radius of the sdO independently from the light curve modeling to test the solution and verify the larger than usual sdO radius, using the atmospheric solution from \S~\ref{section_analysis_spectra} and publicly available distance and photometric data. \textit{Gaia} data release 2 (DR2; \citealt{2018A&A...616A...1G}) allows access to accurate parallax ($\varpi_{\text{Gaia}}$), and thus distance ($d_{\text{Gaia}}$) measurements for $>$1.3 billion stars, including EVR-CB-004 ($\varpi_{\text{Gaia}}=0.4529\pm0.0474$\,mas, $\Delta\varpi_{\text{Gaia}}/\varpi_{\text{Gaia}}\,\lesssim\,0.105$). The combination of $\varpi_{\text{Gaia}}$, surface gravity $g$, effective temperature $T_{\text{eff}}$, and stellar angular diameter $\theta$ allowed us to determine the fundamental stellar parameters, including the radius $R$, mass $M$, and luminosity $\log{L/L_{\odot}}$, of the primary independently. This is referred to as the Spectrophotometric Approach:

\begin{equation}
    R \stackrel{\theta \ll 1}{\approx} d \cdot \frac{\theta}{2} = \frac{\theta}{2\varpi}\label{radius}
\end{equation}
\begin{equation}
    \log(L/L_{\odot}) = \log{(4\pi R^2\sigma T_{\text{eff}}^4/L_{\odot})}\label{luminosity}
\end{equation}
\begin{equation}
    M = \frac{g \theta^2}{4 G \varpi^2}\,,\label{mass}
\end{equation}
where $\sigma$ is the Stefan-Boltzmann constant and $G$ is the Gravitational constant. 
The respective uncertainties are derived from Gaussian error propagation:
\begin{equation}
    \Delta R = \frac{1}{2\varpi}\sqrt{(\Delta\theta)^2 + \theta^2 \cdot \left(\frac{\Delta \varpi}{\varpi}\right)^2}\label{radius uncertainty}
\end{equation}
\begin{equation}
    \Delta \log(L/L_{\sun}) = \frac{2}{\ln{10}}\sqrt{\left(\frac{\Delta R}{R}\right)^2 + 4 \cdot \left(\frac{\Delta T_{\text{eff}}}{T_{\text{eff}}}\right)^2}\label{luminosity uncertainty}
\end{equation}
\begin{equation}
    \Delta M = M\sqrt{(\ln{10}\Delta\log{g})^2 + 4\left(\frac{\Delta\theta}{\theta}\right)^2 + \left(\frac{\Delta\varpi}{\varpi}\right)^2}\label{mass uncertainty}
\end{equation}

We decided not to take the \textit{Gaia} DR2 parallax zero point offset into account as also recommended by \citet{2018A&A...616A...2L} and \citet{2018A&A...616A..17A}, since it depends on the types of astrophysical objects investigated and is still under debate (see, for instance, the different results of \citealt{2018A&A...616A...2L}, \citealt{2018ApJ...861..126R}, \citealt{2019ApJ...878..136Z}, or \citealt{2019MNRAS.487.3568S}). Furthermore, the zero point offset is a function of the coordinates since it depends on \textit{Gaia's} scanning pattern \citep{2018A&A...616A..17A}, which makes it even more difficult to correct for it. Last but not least, we decided not to correct for possible small-scale variations for the parallax measurements, since it is almost impossible to determine them for a single object like EVR-CB-004 \citep{2018A&A...616A...2L}.

The necessary atmospheric parameters ($T_{\text{eff}}$, $\log{g}$) have already been determined in \S~\ref{section_analysis_spectra}. Based on $T_{\text{eff}}$, $\log{(g)}$, and $\log{n(\text{He})}$, the stellar angular diameter $\theta$ can be derived from a spectral energy distribution (SED) fit to appropriate photometric data according to the analysis methodology presented \textbf{by} \citet{2018OAst...27...35H}.

We made use of the following photometric data available on VizieR\footnote{https://vizier.u-strasbg.fr/viz-bin/VizieR}: SkyMapper DR1 \citep{Wolf_2018}, \textit{Gaia} DR2 \citep{2018A&A...616A...1G}, SDSS DR9 \citep{2012ApJS..203...21A}, PanSTARRS DR1 \citep{Chambers_2016}, 2MASS \citep{Skrutskie_2006}, and AllWISE \citep{Wright_2010, Cutri_2013}. All magnitudes used are listed in Table \ref{phot} in the Appendix.

The objective $\chi^2$-based SED fit was carried out within the ``Interactive Spectral Interpretation System'' ISIS, which was designed at the Massachusetts Institute of Technology (MIT) by \citet{2000ASPC..216..591H}. We used two free fit parameters. The stellar angular diameter $\theta$ has the effect of shifting the SED up and down according to $f(\lambda)=[\theta^2 F(\lambda)]/4$, where $F(\lambda)$ is the synthetic model flux at the stellar surface and $f(\lambda)$ is the observed flux at the detector position, whereas the monochromatic color excess $E(44-55)$, based on the monochromatic magnitudes at wavelengths $\lambda=$\,4400\,\AA\,and $\lambda=$\,5500\,\AA, reddens the spectrum. We treated the interstellar extinction via $A(\lambda)$, describing the interstellar extinction in magnitude at wavelength $\lambda$ according to Eq. (1) in \citet{2019ApJ...886..108F}. $A(\lambda)$ can also be expressed in terms of $E(44-55)$ and the extinction coefficient $R(55):=A(5500)/[E(44-55)]$ (see Eqs. (2), (3), and (8) as well as Table 3 in \citealt{2019ApJ...886..108F}). In our case, we fixed $R(55)$ to 3.02, the value for the diffuse interstellar medium in the Milky Way. For the high effective temperature in question the monochromatic reddening parameter is identical to that in the Johnson system \citep[E(B-V), see Table 4 of][]{2019ApJ...886..108F} and the result (E(44-55)=$0.143\pm0.007$\,mag) is consistent with reddening maps of \citet{1998ApJ...500..525S} and \citet{2011ApJ...737..103S}, which give $0.149\pm0.004$\,mag and $0.129\pm0.003$\,mag, respectively.

We rescaled all uncertainties to guarantee a best fit of $\chi_{\text{red}}^2\sim1$. 1$\sigma$ single confidence intervals for $\theta$ and $E(B-V)$ were calculated in the following way: Starting from the best fit with $\chi_{\text{red}}^2\sim1$, we increased/decreased the parameter under consideration, while fitting the other one, until a certain increment $\Delta\chi^2$ from the minimum $\chi^2$ was reached. Chosen values for $\Delta\chi^2$ determined the confidence level of the resulting interval, for instance, $\Delta\chi^2=1$ yielded 1$\sigma$ single confidence intervals.

\begin{figure}[ht]
\includegraphics[width=1.0\columnwidth]{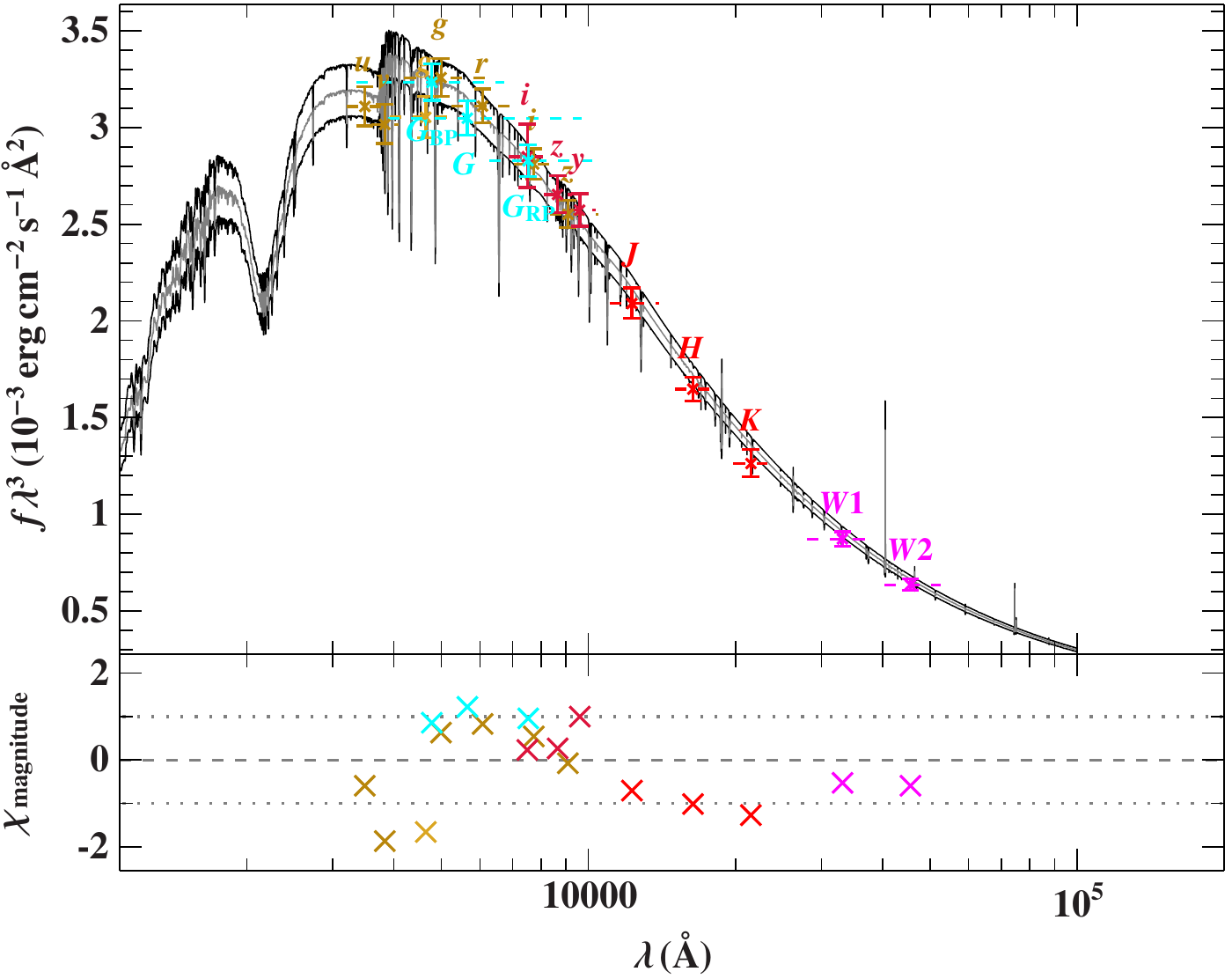}
\caption{Comparison of a synthetic spectrum with photometric data for EVR-CB-004. Filter-averaged fluxes are shown as colored data points that were converted from observed magnitudes (the dashed horizontal lines indicate the respective filter widths). The gray solid line represents a synthetic spectrum based on the final atmospheric parameters derived from the low and medium-resolution SOAR spectra (see Table \ref{solutions_table}), whereas the black solid lines are based on the final values of $\log{(g)}=4.553$ and $\log{n(\text{He})}=-0.843$, but different values of $T_{\text{eff}}=$\,39769\,K and $T_{\text{eff}}=$\,42263\,K, showing the effect of the statistical and systematic uncertainties on $T_{\text{eff}}$ (see Sect. \ref{section_analysis_spectra} for details) on the spectral energy distribution. The panel at the bottom shows the differences between synthetic and observed magnitudes. The following color codes are used to identify the photometric filter systems: SkyMapper and SDSS (yellow), Gaia (cyan), PanSTARRS and 2MASS (red), and AllWISE (magenta). The flux density times the wavelength to the power of three ($f_\lambda \lambda^3$) as a function of wavelength is plotted in order to eliminate the steep slope of the constructed SED over the displayed broad wavelength range.}
\label{SED}
\end{figure}

Figure \ref{SED} shows the resulting SED. Thanks to the very precise photometric data, the uncertainty on the angular diameter ($\Delta\theta$/$\theta$) is 1.6\,\% only. This includes the statistical and systematic uncertainties on $T_{\text{eff}}$ as discussed in Sect. \ref{section_analysis_spectra}, which propagate into the predicted fluxes and, hence, into $\theta$. In conclusion, the mass uncertainty is dominated by the surface gravity uncertainty and the parallax measurement. 

\begin{table}
\caption{Parallaxes and fundamental stellar parameters for the primary of EVR-CB-004 derived from the spectrophotometric approach. Gaia: Based on measured \textit{Gaia} parallax. BJ: Based on distance derived from Bayesian methods \citep{2018AJ....156...58B}. $^\dagger$: 1$\sigma$ statistical uncertainties only. $^\ast$: Listed uncertainties result from statistical and systematic errors (see Sects. \ref{section_analysis_spectra} and \ref{The Spectrophotometric Approach} for details).}\label{photometric results}
\centering
\begin{tabular}{ccc}
\hline\hline
Parameter & Unit & Result\\
\hline
$\varpi_{\text{Gaia}}$ & [mas] & $0.4529\pm0.0474^\dagger$ \\
$d_{\text{Gaia}}$ & [pc] & $2207.993\pm231.086^\dagger$ \\
$\varpi_{\text{BJ}}$ & [mas] & $0.4847^{+0.0538}_{-0.0443}$$^\dagger$ \\
$d_{\text{BJ}}$ & [pc] & $2063.199^{+228.882}_{-188.519}$$^\dagger$ \\
$\theta$ & [$10^{-11}$\,rad] & $1.242\pm0.012^\dagger$ \\
$E(44-55)$ & [mag] & $0.143\pm0.007^\dagger$ \\
$R_{\text{Gaia}}$ & [$R_{\odot}$] & $0.61\pm0.07^\ast$ \\
$M_{\text{Gaia}}$ & [$M_{\odot}$] & $0.48\pm0.13^\ast$ \\
$\log{(L_{\text{Gaia}}/L_{\odot})}$ & & $2.97\pm0.11^\ast$ \\
$R_{\text{BJ}}$ & [$R_{\odot}$] & $0.57^{+0.07}_{-0.06}$$^\ast$ \\
$M_{\text{BJ}}$ & [$M_{\odot}$] & $0.42^{+0.12}_{-0.11}$$^\ast$ \\
$\log{(L_{\text{BJ}}/L_{\odot})}$ & & $2.91^{+0.12}_{-0.10}$$^\ast$ \\
\hline
\label{solutions_table_spma}
\end{tabular}
\end{table}

Table \ref{photometric results} summarizes the spectrophotometric results based on \textit{Gaia}. The given uncertainties on the fundamental stellar parameters result from Eqs. (\ref{radius uncertainty}), (\ref{luminosity uncertainty}), and (\ref{mass uncertainty}), whereby we used the 1$\sigma$ statistical and systematic errors for T$_{\text{eff}}$ and $\log{(g)}$ from \S~\ref{section_analysis_spectra}, and $\Delta\theta$/$\theta\sim$\,1.6\,\%.

We also determined the fundamental stellar parameters from distances derived from Bayesian methods. We used the distance from \citet{2018AJ....156...58B}, converted it to the parallax space via the usual relationship $d=1/\varpi$ and again determined $R$, $M$, and $\log{(L/L_{\sun})}$ via the Spectrophotometric Approach. The results based on Bailer-Jones and the ones derived from \textit{Gaia} are in good agreement (see Table \ref{photometric results}). Both results are also consistent with the light curve modelling. 

\begin{figure*}[ht]
\centering
\includegraphics[width=0.8\textwidth]{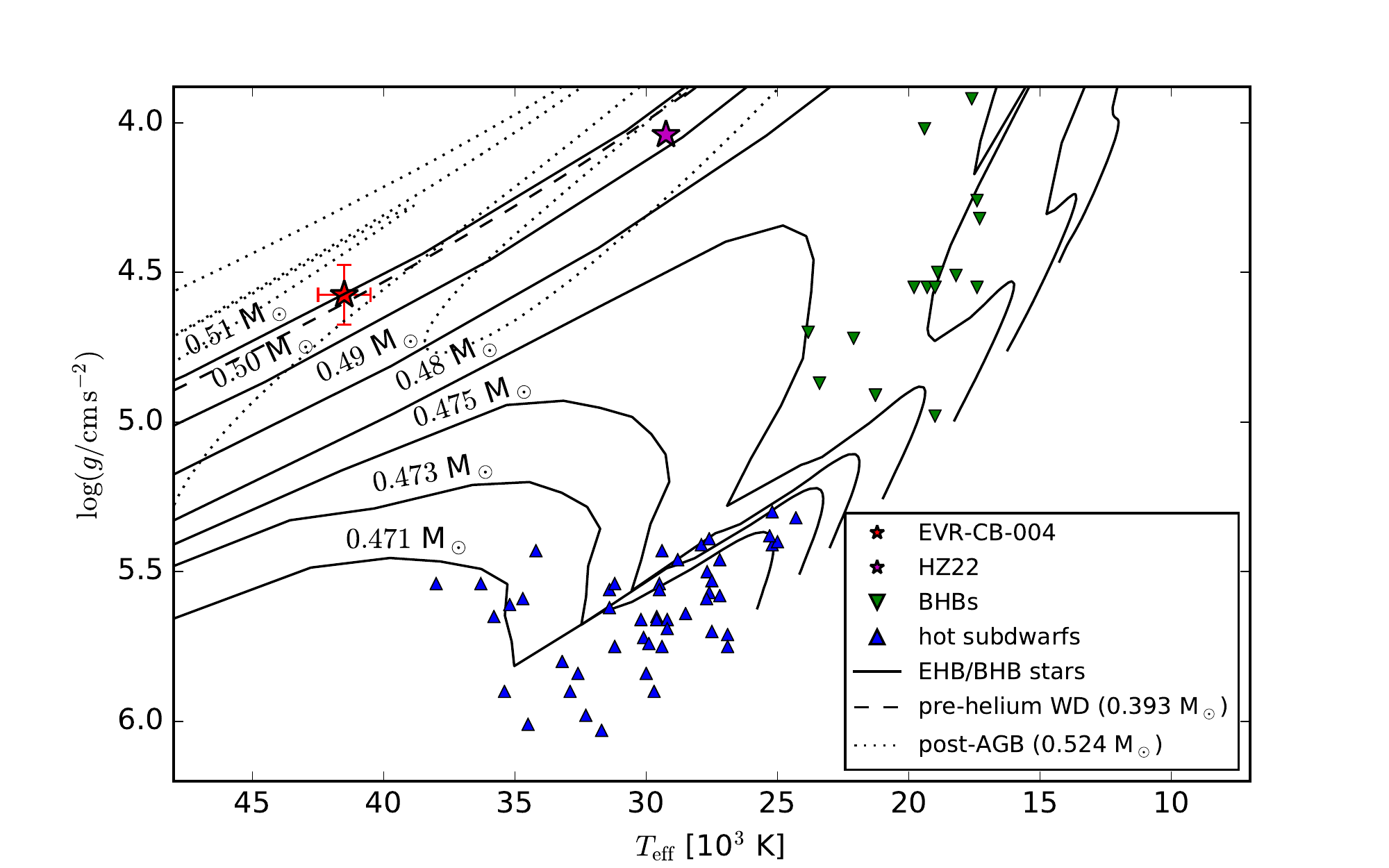}
\caption{$T_{\text{eff}}$-$\log{(g)}$ diagram of the primary star in EVR-CB-004 (red star). EHB/BHB evolutionary tracks for different stellar masses (bottom to top: 0.471\,$M_{\odot}$, 0.473\,$M_{\odot}$, 0.475\,$M_{\odot}$, 0.480\,$M_{\odot}$, 0.490\,$M_{\odot}$, 0.500\,$M_{\odot}$, and 0.510\,$M_{\odot}$), that is, increasing hydrogen envelope mass (0.000, 0.002, 0.004, 0.009, 0.019, 0.029, and 0.039\,$M_{\odot}$, respectively), and solar metallicity according to \citet{1993ApJ...419..596D} are shown with solid lines. In addition, the post-AGB tracks according for 0.524\,$M_{\odot}$, are displayed with dotted lines \citet{1995A&A...299..755B} as well as the pre-helium WD track calculated with MESA \citep{pax11,pax13,pax15,pax18,pax19} shown with dashed lines. The hot subdwarfs are confirmed binaries with WD companions taken from \citet{kup15}. BHB stars are taken from \citet{1997ApJ...491..172S,2001A&A...379..235R,2007ApJ...668L..59V,2010MNRAS.409..582N,2011MNRAS.415.1381C,2012ApJ...753L..17O,2014A&A...562A..95G,2018A&A...618A..86S}. Plotted error bars include 1$\sigma$ statistical and systematic uncertainties as presented in the text (see Sect. \ref{section_analysis_spectra} for details).}
\label{Kiel diagram}
\end{figure*}

\subsection{Unexpected properties of the sdO in EVR-CB-004} \label{section_our_odd_star}

From the spectroscopic data, we classify the primary star in EVR-CB-004 as a hot subdwarf O-type star, and the derived mass is consistent with known hot subdwarfs. However, the sdO is shown to have a considerably lower surface gravity ($\log{g}=4.55$) than expected for a standard shell-burning sdO hot subdwarf (typically $\log{g}=5.5-6.0$ see \citealt{2009CoAst.159...75O}), with a corresponding large radius of $0.6R_{\odot}$. These properties also drive the exceptionally large amplitude (16.0\% change in brightness from maximum to minimum) ellipsoidal modulations. These properties were confirmed independently from the atmospheric and light curve solutions (see the previous sections).

While these values are non-canonical for an sdO, the larger spread in sdO properties indicates this could be a peculiar (inflated) sdO especially considering the mass, temperature, and compact binary system characteristics are consistent with a canonical sdO primary. We rule out this interpretation completely because additional spectral analysis revealed the system to be $\approx$10-100 times more luminous than expected for a shell-burning sdO. Despite its small size and mass, EVR-CB-004 is $\approx$1000 times the solar luminosity. We show in the following section the primary is more likely an evolved hot subdwarf, an unexpected find in an already rare compact binary system. 

Surprisingly, the sdO is close to filling its Roche Lobe or perhaps even fills its Roche Lobe. We would instead expect a post-CE compact binary with a canonical like hot subdwarf to stabilize at a close separation, but beyond any mass transfer point. The fact that the sdO is Roche Lobe filling is novel and may suggest that it has expanded since emerging from the CE in which its progenitor was formed. It is unclear if the system is actively accreting, a possibility given the sdO is so close to filling its Roche Lobe.

These surprising properties must be taken into account when considering the sdO and evolutionary history of the EVR-CB-004 system.

\subsection{Comparison to Stellar Evolution Models} \label{section_nature_of_hsd}

To investigate the nature of the primary, we compare evolutionary tracks of hot subdwarf models of compact pre-helium white dwarfs (pre-He WDs), helium-burning stars and post-asymptotic-branch (post-AGB) stars with our observed properties \citep{1993ApJ...419..596D, 1995A&A...299..755B}. From kinematic analysis we find that EVR-CB-004 is likely a member of the young Galactic thin disc population, see the Appendix for additional details. We note here that for all stellar evolution models, we adopt a solar metallicity, justified by the helium content of the sdO (see the previous section) and population type. Following, we discuss three different interpretations as to the nature of the primary in EVR-CB-004.

In the first scenario, if the progenitor filled its Roche Lobe before reaching the tip of the red giant branch, the star would evolve into a pre-He WD and contract to become a He WD. The mass of the He WD depends on the mass of the helium core when the progenitor filled its Roche Lobe. We use the stellar evolution code \texttt{MESA} \citep{pax11,pax13,pax15,pax18,pax19} to calculate tracks for different pre-He WD models and find that a pre-He WD with a mass of $0.393$\,$M_\odot$ is consistent with the observed T$_{\text{eff}}$ and $\log{(g)}$ (see Fig.\,\ref{Kiel diagram}. This mass is inconsistent with the derived mass from the light curve modelling and because the sdO is close to Roche Lobe filling the pre-He WD would just have been born - therefore we consider this solution to be unrealistic.

If instead the progenitor filled its Roche lobe after He-burning started, the envelope will get stripped and forms a He core-burning hot subdwarf star which is expected to burn He for $\approx100-150$\,Myrs. Depending on its hydrogen envelope it is considered to be extreme horizontal branch (EHB) star (for hydrogen envelopes $\lesssim0.01$\,M$_\odot$) or blue horizontal branch (BHB) star (for hydrogen envelopes of a few hundreds\,M$_\odot$). Once burning exhausts He in the core, the star evolves toward hotter temperatures. As the core contracts, residual hydrogen is predicted to burn in a shell, pushing the surface to a larger radius \citep{2009CoAst.159...75O} and hence lower surface gravities. This is seen in the solid tracks shown for different masses in Figure~\ref{Kiel diagram}. This stage of the evolution is expected to last for only $\approx10-20$ million years and is commonly referred to as post-EHB or post-BHB evolution. A helpful discussion of EHB/BHB stars and their evolution can be found in \citet{2001PASP..113.1162M}, \citet{2009CoAst.159...75O} and \citet{2016PASP..128h2001H}. Figure \ref{Kiel diagram} shows the position of the primary of EVR-CB-004 in the $T_{\text{eff}}$-$\log{(g)}$ diagram, which is consistent with a post-BHB sequence with a mass of $\approx0.5\,M_\odot$. We note here the limitation of using the EHB/BHB evolutionary models in their current form to describe the primary of EVR-CB-004, since their radii exceed the Roche radius of EVR-CB-004 between leaving the BHB and the present. We suggest more involved modeling, beyond the scope of this work, to explore the post-BHB space specific to EVR-CB-004.

If the post-BHB interpretation is correct, the primary of EVR-CB-004 is even rarer as we would have to have caught the object during this transitioning state. The only other reasonably similar system (compact binary with a WD companion, ellipsoidal deformation, Doppler boosting, gravitational limb darkening, similar mass, and an old evolved primary) we found in the literature is HZ 22 \citep{1972ApJ...174...27Y}. However this interesting object is quite different in other ways, with a lower temperature and surface gravity as well as a larger radius.

In addition to post-EHB/BHB evolutionary tracks, we also compared the primary of EVR-CB-004 to post-AGB (post-asymptotic giant branch) tracks \citep{1995A&A...299..755B}). Post-AGBs are also final stage objects transitioning to a WD; an excellent review of post-AGB stars can be found in \cite{2003ARA&A..41..391V}. For a recent survey (of hot UV-bright stars in globular clusters) yielding several post-AGB discoveries along with their atmospheric properties, see \cite{2019A&A...627A..34M}. In Figure~\ref{Kiel diagram}, the post-AGB evolutionary tracks for a slightly more massive object fit our observed values, but there is a significant difficulty with this interpretation. Finding a short lived post-AGB star in a tight binary seems highly unlikely. The post-AGB phase at this part of the observed $T_{\text{eff}}$-$\log{(g)}$ is expected to be fast, on the order of $10^5$ years \citep{2003ARA&A..41..391V}. Because the sdO is close to Roche lobe-filling we would expect that the sdO only recently left the common envelope which is very unlikely due to the short timescale.

We point out the post-AGB discussion above assumes a clean object without a significant remaining CE. It is possible the central star is a post-AGB with a planetary nebula, and would belong to the class of Planetary Nebula binary Central Stars (CSPN), a small subset of which have post-AGB primaries. However the post-AGB models predict a massive CE, while the post-BHB scenario predict a low mass CE of around 0.01 $M_{\odot}$ because the envelope mass of the BHB becomes the CE. The sdO in EVR-CB-004 is sufficiently hot and luminous that we would expect to see emission lines typical of a large PN given a CSPN object. The absence of a PN is an important discriminator that favors the post-BHB interpretation.

All three of the discussed scenarios, the pre-He WD, post-BHB and post-AGB explanations, agree with the lower observed surface gravity, but are relatively short-lived phases that are challenging to explain in the already rare compact system. In addition, the pre-He WD mass is not consistent with known or simulated systems. Given the evolutionary timescale which is a factor of 100 slower for post-BHB stars compared to the other scenarios and the lack of a PN, the most likely explanation is the post-BHB interpretation of the primary in EVR-CB-004.

Extensive spectroscopic analysis (very high resolution and comprehensive wavelength coverage beyond the scope of this work) could constrain the atmospheric parameters that may favor one interpretation more strongly. Two examples revealing post-AGB stars can be found in \cite{1986A&A...169..244H} and \cite{2015MNRAS.452.2292C}. We suggest this as future EVR-CB-004 follow-up work.

\subsection{Formation and Evolution}

The position of EVR-CB-004 in the (T$_\textrm{eff}$-$\log\,g$) diagram could be explained by three different evolutionary scenarios (see Fig. \ref{Kiel diagram}), all assuming that the sdO star formed as a result of being stripped of its envelope by the close white dwarf in different stages of progenitor evolution. When stripping occured while the progenitor was on the first giant branch, EVR-CB-004, would be a helium star evolving into a helium white dwarf. If stripping occurred when the progenitor already ascended the AGB, EVR-CB-004 would evolve into a C/O white dwarf.
In both cases it is likely that stripping led to the formation and subsequent ejection of a common envelope. Given the low mass of the sdO star, the mass of the ejected envelope is likely to be as large as a few tenth of the mass of the sun. Because EVR-CB-004 just came out of its Roche lobe, the envelope should still be detectable. EVR-CB-004 is hot and luminous enough to ionise the ejected material, which means  it should show up as a planetary nebula PNe. A significant fraction of (mostly bipolar PNe are known to host close binary central stars \citep[see][for a review]{2017NatAs...1E.117J} with orbital periods similar to that of EVR-CB-004. In most of those binary central stars the secondary is not a white dwarf, but a late type main sequence star, which can be detected from light curves by the strong reflection effect. It is much more difficult to detect a white dwarf companion, because there would be very little reflected light. Instead the white dwarf would reveal itself by the ellipsoidal light variation of the visible star as observed for EVR-CB-004. Usually, ellipsoidal light variations are small and hard to detect, which produces observational bias against detecting white dwarf companions. Nevertheless, white dwarf companions to central stars have been discovered for a few central stars, e.g. the binary central star of NGC 6026 \citep{2010AJ....140..319H}. These scenarios could be appropriate for EVR-CB-004, if the ejected envelope could be identified as a planetary nebula, which, however, is not the case.

Hence, we consider a third option, that is stripping at the tip of the first giant branch. In this case a hot helium burning star would form as an extreme horizontal branch star. Such stars have been identified as sdB stars \citep{1986A&A...155...33H}. Because they burn helium for more than 100\,Myrs, the ejected envelope has long been dispersed when core helium burning stopped. Such stars retain a tiny hydrogen envelope ($<$0.01 $M_{\odot}$) too thin to ignite hydrogen shell burning. The stars will evolve directly into white dwarf. If, however, the envelope mass is slightly larger (0.01--0.02 $M_{\odot}$) would lie on the blue horizontal branch at effective temperatures of 15,000K to 18,000K. When core helium burning ceases, hydrogen shell burning drives the star to expand and reheat thereafter, leading to a hot sdO star, which is more luminous than most sdOs known. In this post-BHB phase the star will fill its Roche lobe and the tiny envelope will be lost by mass transfer or ejection. Even if a CE event occurred, the ejected mass would have been too small to be detectable as a planetary nebula. 
The lack of a PN around EVR-CB-004 favors the latter post-BHB scenario.
Because also the white dwarf primary could originate in a CE event, the EVR-CB-004 binary could have gone through three phases of mass transfer or ejection.

Because the sdO in EVR-CB-004 is so close to Roche Lobe filling, we discuss briefly the likely accretion during the post-BHB stage. The expansion driven by the post-BHB shell burning will push the radius outward to overflow its Roche lobe and start accretion onto the WD companion. As the sdO star accretes onto the WD companion, the sdO will increase in temperature but maintain a constant radius (still consistent with the observed properties of the primary in EVR-CB-004). We would like to emphasize that the current data does not allow us to exclude an accretion disc and ongoing accretion. The sdO in EVR-CB-004 is consistent with a Roche Lobe filling post-BHB star, and with a luminosity of $\log{(L/L_{\odot})}\approx3$ the sdO would outshine an accretion disc in the optical. Additionally, the inclination angle is too small to show any eclipse from an accretion disc. If the sdO in EVR-CB-004 is actively accreting, it is more extreme (longer period/larger sdO star) than the recently discovered ZTF\,J2130, which was found to be an accreting sdO star at 39\,min orbital period where the sdO gets eclipsed by the accretion disc \citep{2020arXiv200201485K}. It is also possible that we see the EVR-CB-004 system as an active accretor in the short post-BHB window. X-ray analysis could confirm the system as an active accretor, and we leave that followup observation and analysis to future work.

The EVR-CB-004 system is expected to evolve into a double-degenerate WD + WD binary (regardless of the sdO, post-BHB, pre-He WD, or post-AGB interpretation). The orbit will then shrink due to gravitational wave radiation, until the period reaches a few minutes in $\approx4$\,Gyrs. As the orbit shrinks to this small separation, the less massive (but larger radius) WD will fill its Roche lobe and transfer mass to the more massive companion WD. What happens next depends on several factors, most importantly the mass ratio and the total mass; a helpful discussion of WD merger evolution can be found in \cite{2012MNRAS.427..190S}. WD merger simulations performed by \cite{2004MNRAS.350..113M} reveal a narrow range for mass fractions ($2/3 < q < 1$, where q is the mass of the donor / the mass of the accretor) where the WDs are expected to merge via unstable direct impact mass transfer. The mass fraction of EVR-CB-004 ($q=.76$) suggests the system will merge to form a 1.2$M_{\odot}$ high mass single WD. Some extraordinary WD merger systems from the ELM survey are presented in \cite{2012ApJ...751..141K} (see Figure 6), with EVR-CB-004 falling in the high-mass-outlier regime and well placed in the merger region. However, such a large combined mass can also lead to a thermonuclear supernova in $\approx\,$4\,Gyr as discussed in detail in \cite{she18,Perets19,Zenati19}.

Double WD systems as producers of higher mass single WDs is an active area of research. A recent investigation of merger rates for high mass WD merger rates can be found in \cite{2019ApJ...886..100C} showing a less than 10\% rate for WD mergers near the total mass of EVR-CB-004. WDs with masses greater than $\approx1M_{\odot}$, regardless of origin, are predicted and observed to be quite rare. \cite{2016MNRAS.461.2100T} shows rates of a few percent or less, in a sample biased toward the higher mass. EVR-CB-004 is a viable candidate double WD merger forming a single high mass WD or a thermonuclear SN\,Ia, making it a quite rare system from this aspect alone.

\subsection{Low Amplitude Light Variation} \label{section_low_amp_var}

\begin{figure}[ht]
\includegraphics[width=1.0\columnwidth]{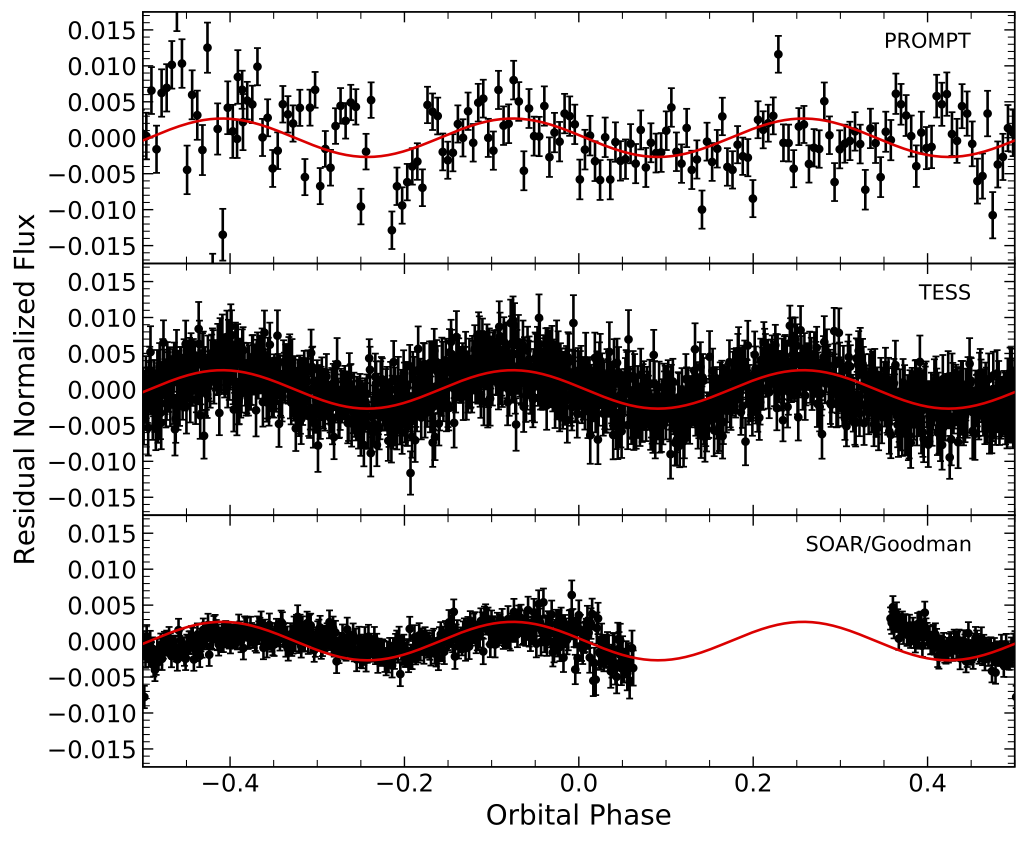}
\caption{Phase-aligned residuals after removing our best--fitting model from the PROMPT (top; R filter), TESS (middle; $\sim$I filter), and SOAR/Goodman (bottom; V filter) light curves. While strongest in the TESS data, all three residuals show hints of an additional variation at one--third the orbital period. The red line shows a simple sinusoidal fit to the TESS residuals, with the period fixed to one--third the orbital period.}
\label{fig:EVR004_lc_residuals}
\end{figure}

In addition to the photometric variations from ellipsoidal deformation, Doppler boosting, and gravitational limb darkening, the high precision SOAR, TESS and PROMPT light curves also show a 2.028 hour low amplitude (0.4\% in TESS) sinusoidal signal. Figure \ref{fig:EVR004_lc_residuals} shows the SOAR, TESS, and PROMPT light curves (phase folded on the 6.084 orbital period), with the residuals after removing the astrophysical signal from the solution in \S~\ref{section_analysis_lc}. Clearly visible in the residuals is a low amplitude signal that is a resonance of the dominant signal. We checked the best period of the residual signals from SOAR and TESS by analysing them with LS and find the results are consistent with the observed period of 2.028 hours. The most challenging aspect of the signal is that the period is a 3:2 ratio with respect to the dominant light curve feature (the ellipsoidal deformation of the primary seen at 3.042 hour cycles) with a phase offset between the low amplitude and dominant light curve signals. This combination of features cannot be due to a poor fit to the data. Following we discuss possible sources of this signal. 

\subsubsection{Asynchronous Rotation}

{\sc Lcurve} assumes that the deformed sdO star is synchronized to the orbit. If the sdO star is rotating faster than synchronization, this could explain an additional light curve signal. However, the low amplitude variability in the SOAR and TESS light curves (and especially the phase offset with the dominant ellipsoidal signal) does not match any potential super-synchronous signal. Additionally, from the spectroscopic fits and our modeling solution, we \textit{do not} see evidence that the sdO is spun-up. The expected rotational velocity from the light curve solution (115 km s$^{-1}$) is very close to the value measured from the spectra (116.5 km s$^{-1}$), and we conclude the sdO is synchronized in rotation with the orbit. We conclude asynchronous rotation does not explain the variability and amplitude, let alone the 2.028 hour resonant period.

\subsubsection{Eccentricity}

{\sc Lcurve} assumes that the system is in a cicular orbit. Therefore, as with asynchronous sdO rotation, an eccentric binary orbit could explain the additional variability. However, the eccentricity would have to be specific to generate a resonant period and symmetric residual pattern. We cannot identify a mechanism to cause this, and it is challenging to explain why it would occur by random chance. 

\subsubsection{Pulsations/Rossby Waves}

Since several classes of pulsating hot subdwarf stars are known, we also consider the possibility that stellar pulsations are the source of the variability. The p-mode V361 Hya (sdBV$_r$) stars exhibit periods on the order of minutes, while the slightly cooler g-mode V1093 Her (sdBV$_s$) stars have longer periods on the order of 45min-2.5 hours; several hybrid pulsators (sdBV$_{rs}$ are known to exist at the  temperature boundary between the two ($\sim$30,000 K). In the above cases, the pulsations are driven by the $\kappa$-mechanism excited by an opacity bump due to iron abundance enhancement \citep{2003ApJ...597..518F}. Two helium-rich hot subdwarf stars, LS IV-14$^{\circ}$116 and Feige 46, show g-mode pulsations (with P$\sim$1 hr) at  hotter temperatures than the V1093 Her stars. Both the $\epsilon$-mechanism and the $\kappa$-mechanism (due to enhanced C/O abundance) have been proposed to explain these stars \citep{mil11, sai19}. Finally, a new class of pulsating stars, the Blue Large-Amplitude Pulsators (BLAPs), was uncovered recently with temperatures and surface gravities similar to sdB stars and pulsation periods from 3-40 min \citep{pie17,kup19}. They are likely also driven by the $\kappa$-mechanism via helium opacities. The combination of EVR-CB-004's log g-T$_{\mathrm eff}$ values and the 2.028-hr period of the modulation make it unlikely this signal can be explained as any of the aforementioned pulsations driven by the $\kappa$-mechanism. The $\epsilon$-mechanism, on the other hand, could be at play, but this would require EVR-CB-004 to have a helium-burning shell (which is possible if the primary is a post--BHB star).

Since the photometric modulation has a frequency {\em exactly} three times the rotational frequency, one likely explanation is that the variation is a global Rossby-wave (r-mode) oscillation \citep{tow03,sai82}. These surface waves can be  excited in the atmospheres of rotating objects, and have  been identified in {\em Kepler} light curves of hundreds of eclipsing binaries \citep{sai19b}. TESS photometry also revealed potential r-mode oscillations in some helium-rich hot subdwarf stars \citep{jef20}. No other mechanism would predict photometric variations {\em only} at integer factors of the rotational/orbital frequency.

\subsubsection{Source Field}

\begin{figure}[ht]
\includegraphics[width=1.0\columnwidth]{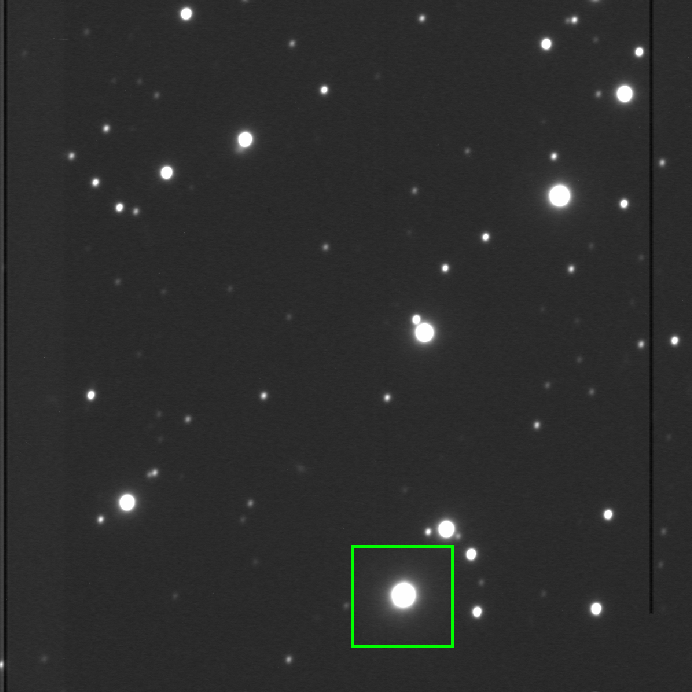}
\caption{The EVR-CB-004 field as seen from stacking the 515 SOAR 20 second images (in V band) to form this final deep image. EVR-CB-004 is the brightest star in the image, located near the bottom center. There are no signs of nebula near the source. The green box is one TESS pixel, with the nearby sources to the right and upper right being potentially blended in the TESS aperture photometry. From the SOAR data, we verified these sources are non-variable and minor in flux (2.5$\%$) compared to the target. The consistent light curve solutions from the SOAR, TESS, and PROMPT data also shows these sources are inconsequential in the TESS data. The image is 3' x 3'.}
\label{fig:EVR004_SOAR_stacked}
\end{figure}

The EVR-CB-004 field has several dim stars near the target, easily separated in the SOAR high resolution images. To check for possible blending in the TESS field and to look for signs of nebula around the target, we stack the 515 SOAR 20 second images to form the deep image of the field (shown in Figure \ref{fig:EVR004_SOAR_stacked}). EVR-CB-004 is the brightest star in the field, near the bottom center. The star to the upper right of EVR-CB-004 and the two dimmer stars to the right are not blended in the TESS pixels, and the other very dim sources nearby are inconsequential (they look exaggerated since this is a 3-hour image from a 4.3m telescope). However the three nearby stars could still contaminate the TESS aperture photometry, which we check in several ways described below.

The crowded field leads to two concerns - influencing the best fit from the light curve solution, and potentially adding an additional variability source. To address the first concern, we fit both the SOAR and TESS light curves independently and the solutions converged on the same results within the reported error ranges. We also adjusted the TESS light curve, based on measurements of the nearby stars using the SOAR data, and found the effect to be minimal and to have no measurable change in our solution. 

To address the concern of added variability, we extracted light curves for each of the potential contaminant stars with the same photometric pipeline used to make the SOAR light curve for EVR-CB-004. We measured the combined contribution of the three potential TESS contaminant sources to be 2.5\%, and we also confirmed they are non-variable. Figure \ref{fig:SOAR contaims} in the appendix shows the light curve of these nearby sources folded on the orbital and on the 1:3 alias periods.

The PROMPT data also provides an opportunity to test the potential contaminant stars. Here we extracted light curves for each of the nearby stars with the same photometric pipeline used to make the PROMPT light curve for EVR-CB-004. Figure \ref{fig:PROMPT_contaims} in the appendix shows the light curve of these nearby sources folded on the orbital and on the 1:3 alias periods, and confirms the non-variability of the SOAR analysis. In the R passband of the PROMPT data, the combined contribution from the nearby stars increases to 35\% of the total flux of the target. With this level of contamination, the amplitude of the main variability would be diluted in the TESS light curve and would influence the system solution. Since our system solution is consistent through all light curves, we conclude the nearby stars do not contribute to the TESS photometry in any significant way.

\begin{figure}[ht]
\includegraphics[width=1.0\columnwidth]{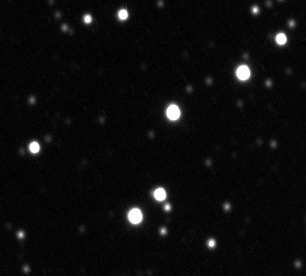}
\caption{The EVR-CB-004 field as seen from stacking the 180 PROMPT 2 minute images (in R band) to form this final deep image. EVR-CB-004 is the brightest star in the image, located near the bottom center. Consistent with the SOAR deep field image, there are no signs of nebulosity near the source. The image is 3' x 3'.}
\label{fig:EVR004_PROMPT_stacked}
\end{figure}

With the deep image, we check for any signs of nebula surrounding the source as this could lead to an additional light curve variation. The PROMPT data is also stacked to form a deep image in R band, and is shown in Figure \ref{fig:EVR004_PROMPT_stacked}. There is no evidence of nebula and we conclude this is not a contributing factor to the low amplitude light curve variation.

\subsubsection{Calcium Lines}

H and K lines of calcium are visible in the SOAR medium resolution RV spectra, which could be indicative of debris or accretion. They are not visible in the low resolution spectra as the resolution is too low to detect the features, and they are not visible in the CHIRON data because the wavelength coverage is beyond the 3933\AA\ and 3968\AA\ CaK and CaH absorption lines. The calcium lines are stable in radial velocity and in amplitude within our measurement uncertainty, and we conclude they do not emanate from the EVR-CB-004 system and are most likely interstellar.

\subsubsection{Unexplained Source}

We have considered all of the obvious (to us) potential sources of the 1/3rd period variability, even including some quite speculative in nature. We acknowledge there could be an astrophysical source we have not thought of that drives this low-amplitude signal. To understand a potential unexplained source, we briefly discuss the approach used in modeling ellipsoidal variable stars.

It is convenient and effective to use a cosine series to analyze ellipsoidal variable star light curves, with the argument being a function of the frequency of the binary orbit. The second harmonic dominates, however the third harmonic is still significant. Higher order terms are inconsequential and are neglected. The amplitudes depend primarily on the radii, mass ratio, orbital inclination, and darkening coefficients. A very good explanation of this approach can be found in \cite{morris1985}, with the same methodology used in the {\sc lcurve} algorithm we employed in \S~\ref{section_analysis_lc} to solve the EVR-CB-004 system.

The models fix the phases of the harmonic terms in order to fit the standard ellipsoidal distortion. The low-amplitude signal in the EVR-CB-004 is not in phase with the main light curve variability, and likely the third harmonic term in {\sc lcurve} does not capture the full variability as well as it is intended due to this phase offset. It could be possible some source of asynchronism is responsible. This partially drove us to consider the many different explanations explored in this section.

\subsubsection{Preferred Solution}

Each of the potential solutions to the low amplitude oscillations has challenges, and we have eliminated to our satisfaction all but the asynchronous, unexplained source, or pulsation options. We note that the asynchronous rotation is the simplest explanation, however the measured rotational velocity does not support this conclusion. An unexplained source is certainly possible, but this is limited to speculation. This leads us to favor the pulsator explanation, with the acknowledgement that additional followup is needed to definitively confirm. Although beyond the scope of this work, extremely high precision multi-band photometric analysis and time series spectroscopy (as performed in the followup works \citealt{2007A&A...471..605V, 2010MNRAS.403..324B, 2019ApJ...878L..35K}) could reveal phase dependent variations in velocity, $T_{\text{eff}}$, and $\log{(g)}$ matching the 2.028 hour light curve low amplitude oscillations. 


\section{SUMMARY} \label{section_summary}

We present the discovery of EVR-CB-004 --- a new 6.08-hr compact binary with a remnant core primary and unseen white dwarf companion. The primary is similar in mass and temperature, 0.52 $M_{\odot}$ and 41,250 K, to an sdO hot subdwarf. However the inflated radius and lower surface gravity of 0.63 $R_{\odot}$ and 4.55 $\log{g}$ suggest a more evolved object.

Our analysis in \S~\ref{section_nature_of_hsd} shows the primary in EVR-CB-004 is likely a more evolved hot subdwarf, possibly caught during its transition from a core He-fusing BHB star to a WD.  The post-BHB stage of hot subdwarf evolution is not well understood, with a limited number of examples to test and verify theoretical models. Finding a post-BHB in a compact binary with a WD is very suggestive that this evolutionary theory is correct, however none have been found. Although additional followup is needed to definitively confirm the primary in EVR-CB-004 as a post-BHB, the evidence from our discovery and followup is strong. The mass and high luminosity are both consistent with a shell-fusing post-BHB star that evolved from a core He burning BHB object. The radius, surface gravity, and high temperature are all in agreement with post-BHB model tracks, but we note such tracks have limited use here as their predicted radii between the BHB stage and EVR-CB-004's current state exceed the Roche radius. Nonetheless, the EVR-CB-004 system is the first viable candidate for a post-BHB + WD compact binary, and with the advantageous characteristics that allow for a complete and precise solution. This includes high amplitude and multiple component variability in the light curve, large radial velocity variations, a robust spectrum with many well resolved features, and bright apparent magnitude. It offers an excellent opportunity to study late-stage hot subdwarf evolution theory and compact binary models.

Besides the post-BHB and rare compact binary, EVR- CB-004 revealed other surprising features. The primary star in EVR-CB-004 is very close to filling its Roche Lobe and thus the system might be actively accreting. We suggest X-ray follow--up observations to confirm and measure any such accretion. The final evolutionary state of the system is also intriguing. EVR-CB-004 is expected to first form a WD + WD binary once the post-BHB and final WD contraction phases complete; it will then likely merge into a very-high mass single WD or a double-detonation under-luminous supernova in $\approx$ 4 Gyrs. Not surprisingly, progenitors to these final stages are sought after and needed to advance our understanding.

In addition to the ellipsoidal modulation, Doppler boosting, gravity darkening and limb darkening components, the light curve of EVR-CB-004 also shows a completely unexpected sinusoidal variation at the $0.4\%$ level with a period that is a 1/3rd resonance (2.028 hours) of the orbital period. This low-amplitude variation has not been seen before in sdO/sdB + WD compact binaries, and is a surprising feature. In section \S~\ref{section_discussion} we discuss our followup analysis to verify this signal is astrophysical, possible explanations, and our preferred pulsator interpretation.

This object was discovered using Evryscope photometric data in a southern-all-sky hot subdwarf variability survey. The multi-component light curve features (bright 13.1 $m_{g}$ source, large amplitude ellipsoidal modulations, Doppler boosting, and gravitational limb darkening), the remnant primary, large WD companion, additional resonant period variation, and merger candidate are unexpected and make EVR-CB-004 an exciting discovery and a unique system.

\section*{Acknowledgements}

This research was supported by the NSF CAREER grant AST-1555175 and the Research Corporation Scialog grants 23782 and 23822. HC is supported by the NSFGRF grant DGE-1144081. BB is supported by the NSF grant AST-1812874. DS is supported by the Deutsche Forschungsgemeinschaft (DFG) under grant HE 1356/70-1. TRM was supported by STFC grant ST/P000495/1. This work was supported by the National Science Foundation through grant PHY-1748958. We acknowledge the use of the Center for Scientific Computing supported by the California NanoSystems Institute and the Materials Research Science and Engineering Center (MRSEC) at UC Santa Barbara through NSF DMR 1720256 and NSF CNS 1725797. The Evryscope was constructed under NSF/ATI grant AST-1407589. This work made use of the spectral energy distribution fitting routine written by A. Irrgang. We made use of ISIS functions provided by ECAP/Remeis observatory and MIT (\url{http://www.sternwarte.uni-erlangen.de/isis/}). We thank J. E. Davis for the development of the \texttt{slxfig} module, which has been used to prepare figures in this work. \texttt{matplotlib} \citep{2007CSE.....9...90H} and \texttt{NumPy} \citep{2011CSE....13b..22V} were used in order to prepare figures in this work. This work has made use of data from the European Space Agency (ESA) mission {\it Gaia} (\url{https://www.cosmos.esa.int/gaia}), processed by the {\it Gaia}
Data Processing and Analysis Consortium (DPAC, \url{https://www.cosmos.esa.int/web/gaia/dpac/consortium}). Funding for the DPAC has been provided by national institutions, in particular the institutions participating in the {\it Gaia} Multilateral Agreement. We made extensive use of NASAs Astrophysics Data System Bibliographic Service (ADS) and the SIMBAD and VizieR database, operated at CDS, Strasbourg, France.

\clearpage

\appendix \label{section_appendix}

\begin{figure}[h]
\includegraphics[width=1.0\columnwidth]{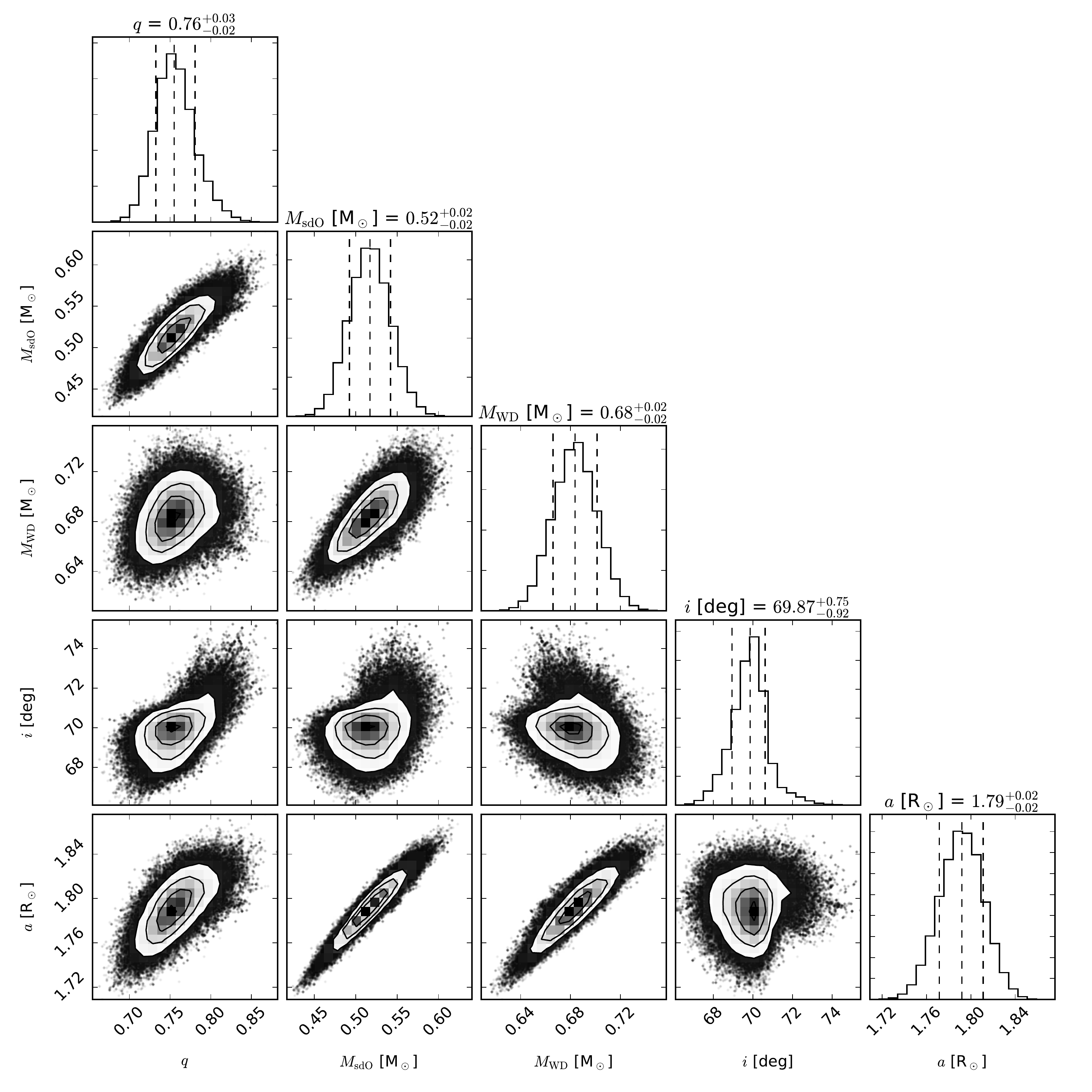}
\caption{Corner plots of the lightcurve fit of EVR-CB-004. The solution converged at masses of $0.68 M_{\odot}$ for the WD and $0.52 M_{\odot}$ for the sdO. The solution prefers an inflated sdO radius of $0.63 R_{\odot}$. The x-axes show, from left to right: $q, M_{\textrm{sdOB}}, M_{\textrm{WD}}$, inclination angle $i$ and separation $a$.}
\label{fig:mcmc_results}
\end{figure}

Figure \ref{fig:mcmc_results} shows the corner plots demonstrating the light curve goodness of fit and convergence. 

\clearpage

Listed in Table \ref{table:sed_fitting_data} is the data used for the SED fitting.

\begin{table}
\centering
	\caption{Photometric data of EVR-CB-004 used for the SED fitting. $^\star$: 1$\sigma$ statistical uncertainties only; $^\dagger$: Extracted from: http://skymapper.anu.edu.au/cone-search/requests/9AVUPMK7/edit/}
	\label{phot}
 \begin{tabular}{ccccc}
 \hline
 System & Passband & Magnitude & Uncertainty & Reference \\
 \hline
 Gaia & G & 13.1266 & 0.0023$^\star$ & (Gaia Collaboration et al. 2018, Gaia DR2: I/345/gaia2) \\
 Gaia & GBP & 12.9693 & 0.0093$^\star$ & (Gaia Collaboration et al. 2018, Gaia DR2: I/345/gaia2) \\
 Gaia & GRP & 13.2841 & 0.0072$^\star$ & (Gaia Collaboration et al. 2018, Gaia DR2: I/345/gaia2) \\
 SDSS & g & 13.0270 & 0.0060$^\star$ & (Ahn et al. 2012, SDSS DR9) \\
 SkyMapper & u & 12.7480 & 0.0030$^\star$ & (Wolf et al. 2018, SkyMapper DR1$^\dagger$) \\
 SkyMapper & v & 12.8860 & 0.0030$^\star$ & (Wolf et al. 2018, SkyMapper DR1$^\dagger$) \\
 SkyMapper & g & 13.0620 & 0.0030$^\star$ & (Wolf et al. 2018, SkyMapper DR1$^\dagger$) \\
 SkyMapper & r & 13.3200 & 0.0030$^\star$ & (Wolf et al. 2018, SkyMapper DR1$^\dagger$) \\
 SkyMapper & i & 13.6760 & 0.0030$^\star$ & (Wolf et al. 2018, SkyMapper DR1$^\dagger$) \\
 SkyMapper & z & 13.9660 & 0.0040$^\star$ & (Wolf et al. 2018, SkyMapper DR1$^\dagger$) \\
 PanSTARRS & i & 13.6400 & 0.0516$^\star$ & (Chambers et al. 2016, PanSTARRS DR1: II/349/ps1) \\
 PanSTARRS & z & 13.8743 & 0.0160$^\star$ & (Chambers et al. 2016, PanSTARRS DR1: II/349/ps1) \\
 PanSTARRS & y & 14.0211 & 0.0060$^\star$ & (Chambers et al. 2016, PanSTARRS DR1: II/349/ps1) \\
 2MASS & H & 13.7170 & 0.0270$^\star$ & (Skrutskie et al. 2006, 2MASS: II/246/out) \\
 2MASS & J & 13.5910 & 0.0270$^\star$ & (Skrutskie et al. 2006, 2MASS: II/246/out) \\
 2MASS & K & 13.8150 & 0.0520$^\star$ & (Skrutskie et al. 2006, 2MASS: II/246/out) \\
 AllWISE & W1 & 13.8210 & 0.0260$^\star$ & (Wright et al. 2010; Cutri et al. 2013, AllWISE: II/328/allwise) \\
 AllWISE & W2 & 13.8750 & 0.0360$^\star$ & (Wright et al. 2010; Cutri et al. 2013, AllWISE: II/328/allwise) \\
 \hline
 \end{tabular}
 \label{table:sed_fitting_data}
 \end{table}

\clearpage

\begin{figure}[h]
\includegraphics[width=0.5\columnwidth]{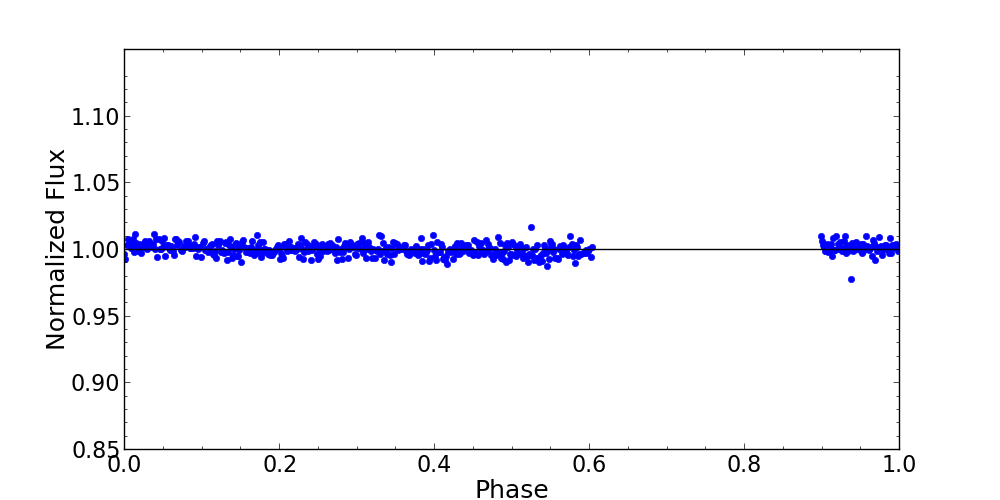}
\includegraphics[width=0.5\columnwidth]{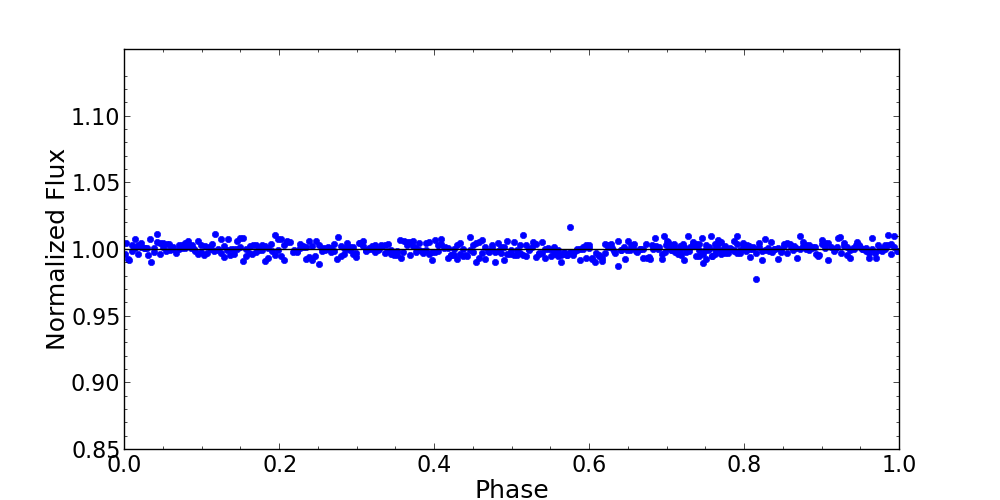}
\caption{\textit{Left:} The combined light curve from the SOAR data of the nearby stars. The data is folded on the 6.084 hour orbital period, and shows no signs of variability. The total flux of these three stars is 2.5$\%$ of the total flux from EVR-CB-004, shown normalized here. \textit{Right:} The same data folded on the 2.028 hour alias period, again showing no signs of variability. This analysis demonstrates the potential contaminants in the TESS photometric aperture do not introduce additional variability into the light curve. Most notably the low amplitude resonant signal cannot be attributed to a TESS blended pixel systematic.}
\label{fig:SOAR contaims}
\end{figure}

Figure \ref{fig:SOAR contaims} shows the combined light curve from the SOAR data of the three nearby stars, processed with the same photometric pipeline used to generate the EVR-CB-004 SOAR light curve. As the sources are potentially blended in the TESS pixels, we check to make sure they do not introduce additional variability into the light curve. They are shown here to be non variable.

\begin{figure}[h]
\includegraphics[width=0.5\columnwidth]{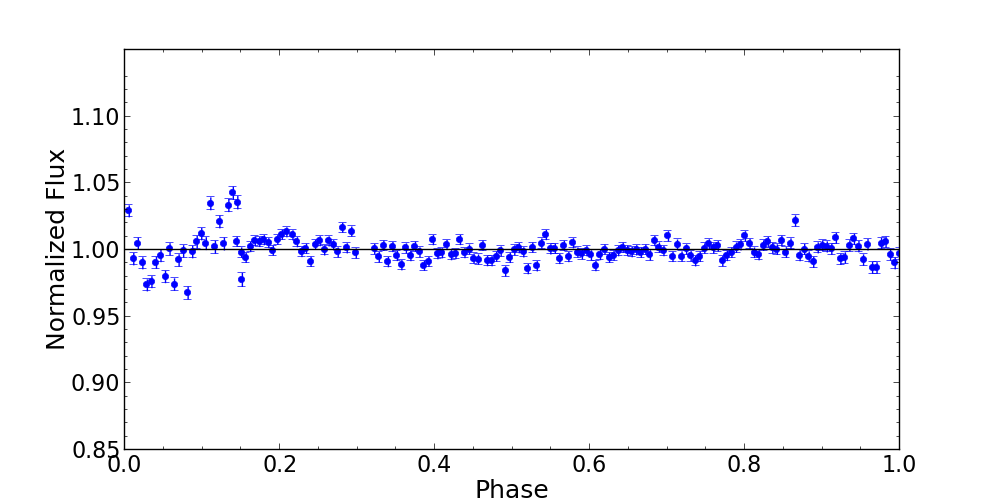}
\includegraphics[width=0.5\columnwidth]{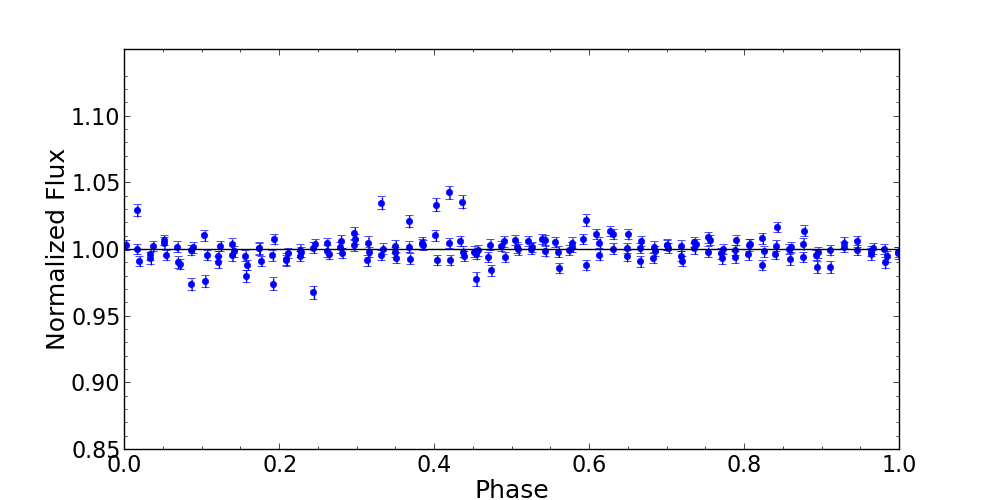}
\caption{\textit{Left:} The combined light curve from the PROMPT data of the three nearby stars. The data is folded on the 6.084 hour orbital period, and shows no signs of variability. \textit{Right:} The same data folded on the 2.028 hour alias period, again showing no signs of variability.}
\label{fig:PROMPT_contaims}
\end{figure}

Figure \ref{fig:PROMPT_contaims} shows the combined light curve from the PROMPT data of the three nearby stars, processed with the same photometric pipeline used to generate the EVR-CB-004 PROMPT light curve. In the PROMPT R passband, the total flux of these three stars increases to 35$\%$ of the total flux from EVR-CB-004. This concern is mitigated by the constant signal that again demonstrates the potential contaminants in the TESS photometric aperture do not introduce additional variability into the light curve. Most notably the low amplitude resonant signal cannot be attributed to a TESS blended pixel systematic. The constant signal in this filter could dilute the EVR-CB-004 light curve amplitude, and consequently affect the fit. The main light curve variation shows no signs of this, the amplitudes are consistent from the different observations, and independent system solutions are the same (within the measurement precision) using SOAR, PROMPT, and TESS data. We therefore conclude the nearby stars did not contribute in any significant way to the TESS photometry.

\clearpage

\begin{figure}
\includegraphics[width=0.5\columnwidth]{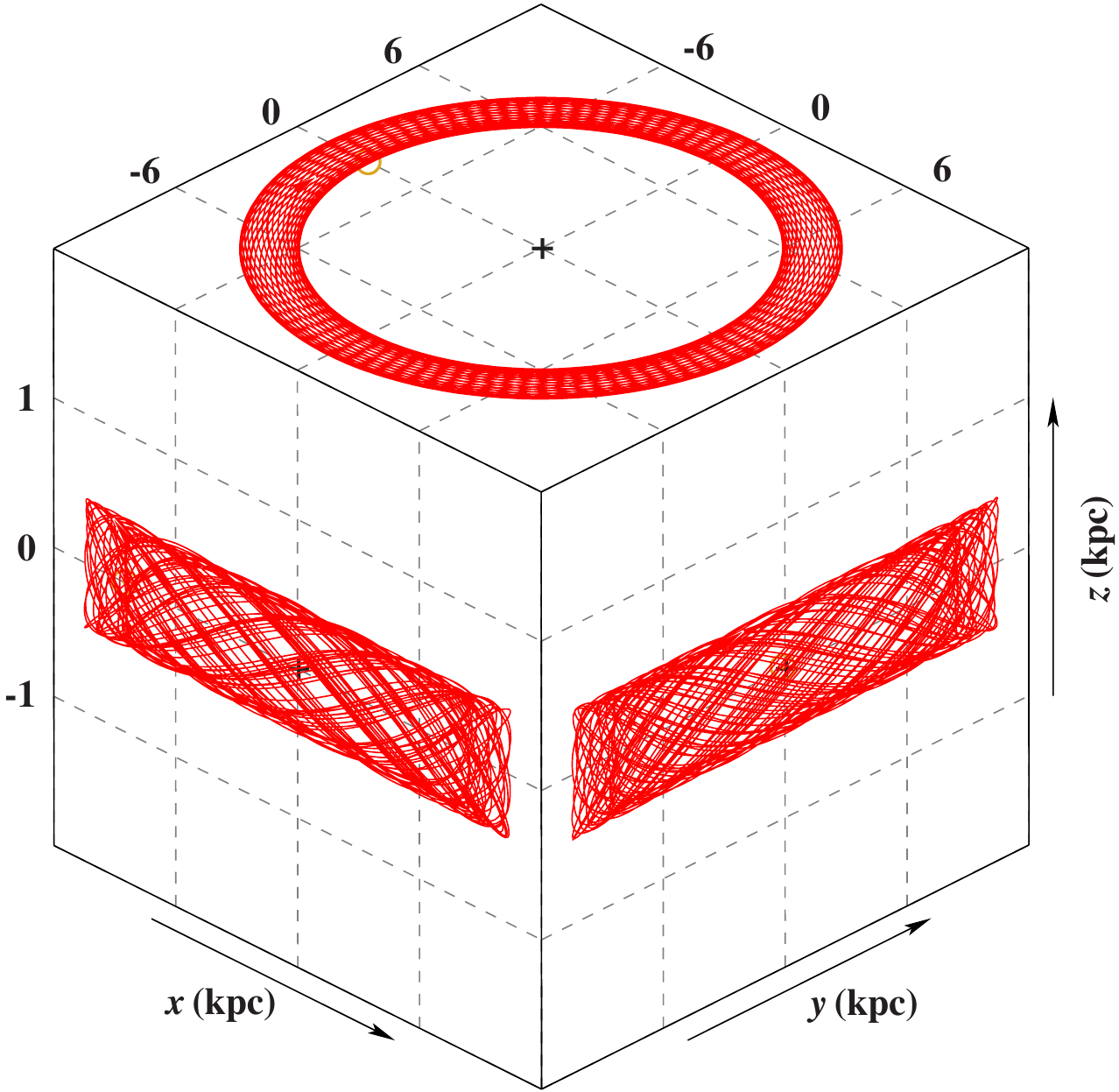}
\caption{\textbf{Galactic orbit of EVR-CB-004 in the Galactic $x$-$y$ and $x$-$z$ planes.}}
\label{fig:galactic_orbit}
\end{figure}

We demonstrate that EVR-CB-004 is likely a member of the Galactic thin disc population by performing a kinematic analysis. We studied the kinematics of EVR-CB-004 by integrating the equation of motion in a Galactic motions using the code developed by
\citet{2013A&A...549A.137I} and the Galactic mass model of \citet{1991RMxAA..22..255A}. The resulting Galactic orbit is shown in Fig. \ref{fig:galactic_orbit}.

In order to study the characteristics of the Galactic orbits we calculated the Galactic velocity components $U$, $V$, and $W$ as described by \citet{2013A&A...549A.137I}, the $z$ component of the orbital angular momentum, and the eccentricity of the Galactic orbit as described in \citet{2003A&A...400..877P} and constructed diagnostic diagrams, that is the $U$-$V$ and $J_z$-$e$ diagram to compare with the kinematical properties of Galactic stellar populations. 

\clearpage
\bibliographystyle{apj}
\bibliography{hsd_wd_binary_2}

\end{document}